\documentclass[copyright,creativecommons]{eptcs}
\providecommand{\event}{ICLP 2026}

\usepackage{iftex}

\ifpdf
  \usepackage{underscore}         
  \usepackage[T1]{fontenc}        
\else
  \usepackage{breakurl}           
\fi

\PassOptionsToPackage{dvipsnames}{xcolor}
\usepackage{tikz}
\usetikzlibrary{positioning, calc}
\usepackage{xspace}
\usepackage{amsmath}

\usepackage{amsthm}
\newtheorem{definition}{Definition}

\newtheorem{example}{Example}
\usepackage{bussproofs}
\EnableBpAbbreviations
\usepackage{pifont}
\usepackage{marvosym}
\usepackage[capitalize]{cleveref}

\Crefname{lstlisting}{Listing}{Listings}

\usepackage{nicematrix}
\usepackage{listings}

\definecolor{lightgreen}{HTML}{558564}

\newcommand\voc{\mathcal{V}\xspace}
\newcommand\theory{\mathcal{T}\xspace}
\newcommand\struct{\mathcal{S}\xspace}
\newcommand\domain{\mathcal{D}\xspace}

\newcommand\imply{\Rightarrow\xspace}
\newcommand\limply{\Leftarrow\xspace}
\newcommand\ifff{\Leftrightarrow\xspace}
\newcommand\fn[2]{#1/#2\xspace}

\newcommand\define{\ensuremath{\overset{\textit{def}}{=}}\;\;}
\newcommand\inputStructure{\struct_{\mathit{in}}}
\newcommand\inputVocabulary{\voc_{\mathit{in}}}

\newcommand\outputVocabulary{\voc_{\mathit{out}}}

\newcommand\pigeon{\text{Pig}\xspace}
\newcommand\hole{\text{Hol}\xspace}
\newcommand\sit{\text{Sit}\xspace}
\newcommand\Node{\text{Node}\xspace}
\newcommand\Edge{\text{Edge}\xspace}
\newcommand\Connected{\text{ToAll}\xspace}

\newcommand\snand{\textsc{snand}\xspace}
\newcommand\snor{\textsc{snor}\xspace}
\newcommand\stn{\textsc{stn}\xspace}
\newcommand\eprop{\textsc{eprop}\xspace}
\newcommand\epred{\textsc{epred}\xspace}
\newcommand\iq{\textsc{iq}\xspace}
\newcommand\trivial{\textsc{trivial}\xspace}
\newcommand\unsat{\textsc{unsat}\xspace}
\newcommand\splitc{\textsc{splitc}\xspace}

\newcommand\pyclingo{\textsc{pyclingo}\xspace}

\newcommand\idpztool{\textsc{IDP-Z3}\xspace}

\newcommand\FOXProb{\langle\voc=\inputVocabulary\cup\outputVocabulary,\inputStructure,\theory\rangle}
\newcommand\FOXProbshort{\langle\voc,\inputStructure,\theory\rangle}

\newcommand{\logicname}[1]{\textsc{#1}\xspace}
\newcommand{\fodot}{\logicname{FO(\ensuremath{\cdot})}} 
\newcommand{\idpz}{\logicname{IDP-Z3}}

\usepackage{xspace}
\usepackage{amssymb}
\usepackage{amsmath}
\usepackage{ifthen}
\usepackage[dvipsnames]{xcolor}
\usepackage{url}

\newcommand\thanksCertiFOX{Funded by the European Union (ERC, CertiFOX, 101122653). Views and opinions expressed are however those of the author(s) only and do not necessarily reflect those of the European Union or the European Research Council. Neither the European Union nor the granting authority can be held responsible for them.} 

\providecommand{\ignore}[1]{}

\newboolean{nocomments}
\setboolean{nocomments}{false}

\newboolean{comments-in-margin}
\setboolean{comments-in-margin}{false}

\newcommand{\namedcomment}[3]{%
	\ifthenelse{\boolean{nocomments}}%
	{\relax}
	{
		\ifthenelse{\boolean{comments-in-margin}}%
		{ \marginpar{\color{#3}{\sc #2} #1}}  
		{  {\color{#3} \textsc{ #2}: #1}  } 
	}%
}

\newcommand{\toolname}[1]{\textsc{#1}\xspace}

\newcommand\trval[1]{\mathbf{#1}}
\newcommand{\ltrue}{\trval{t}}
\newcommand{\lfalse}{\trval{f}}

\newcommand{\geqp}{\geq_{\mathrm{p}}}

\usepackage{etoolbox}

\ExplSyntaxOn

\newcommand\setcitation[2]{%
	\csdef{mycommoncitation\text_uppercase:n{#1}}{#2}%
	\csappto{bbAllCommonCitations}{\cite{#2}\ }
}
\newcommand\getcitation[1]{%
	\csuse{mycommoncitation\text_uppercase:n{#1}}}

\newcommand\refto[1]{%
	\ifcsname  mycommoncitation\text_uppercase:n{#1}\endcsname%
	\getcitation{#1}%
	\else%
	#1%
	\fi%
}

\ExplSyntaxOff

\newcommand\mycite[1]{\cite{\refto{#1}}}

\setcitation{IDP}{DBBJD18PredicatelogicmodelinglanguageIDPsystem}
\setcitation{fodot}{DT08logicnonmonotoneinductivedefinitions}
\setcitation{CPsupport}{DBDD13ModelExpansionPresenceFunctionSymbolsUsing}
\setcitation{minisatid}{DBDD13ModelExpansionPresenceFunctionSymbolsUsing}
\setcitation{FunctionDetection}{DB13Detectionexploitationfunctionaldependenciesmodelgeneration}
\setcitation{lazyGrounding}{DDBS15LazyModelExpansionInterleavingGroundingSearch} 
\setcitation{Bootstrapping}{BJDJBD16BootstrappingInferenceIDPKnowledgeBaseSystem}
\setcitation{BootstrappingIDP}{BJDJBD16BootstrappingInferenceIDPKnowledgeBaseSystem}
\setcitation{FOSymmetry}{DBBD16localdomainsymmetrymodelexpansion}
\setcitation{FOLASP}{VDV21FOLASPFOInputLanguageAnswerSet}
\setcitation{inputster}{JJJ13CompilingInputFOinductivedefinitionsinto}
\setcitation{LearningPaper}{BBBDDJLR15Predicatelogicmodelinglanguagemodelingsolving}

\setcitation{foid}{DT08logicnonmonotoneinductivedefinitions}
\setcitation{cplogic}{VDB10FOIDextensionDLrules}
\setcitation{clog}{BVDV14FOCKnowledgeRepresentationLanguageCausality} 
\setcitation{foc}{BVDV14FOCKnowledgeRepresentationLanguageCausality}
\setcitation{inferenceClog}{BVDV14InferenceFOCModellingLanguage}
\setcitation{inferenceFOC}{BVDV14InferenceFOCModellingLanguage}
\setcitation{examplesClog}{BVDV14FOCRelatedModellingParadigms} 
\setcitation{satid}{MWDB08SATIDSatisfiabilityPropositionalLogicExtendedInductive}

\setcitation{fodot2asp}{DLTV19informalsemanticsAnswerSetProgrammingTarskian} 
\setcitation{Tarskian}{DLTV19informalsemanticsAnswerSetProgrammingTarskian} 
\setcitation{TarskianSemanticsASP}{DLTV19informalsemanticsAnswerSetProgrammingTarskian} 
\setcitation{KBS}{DV08BuildingKnowledgeBaseSystemIntegrationLogic}
\setcitation{KBS-invitedtalk}{D16FOKnowledgeBaseSystemproject}
\setcitation{KBPE}{DWD11PrototypeKnowledge-BasedProgrammingEnvironment}

\setcitation{justifications}{DBS15FormalTheoryJustifications} 
\setcitation{justificationTheory}{DBS15FormalTheoryJustifications} 
\setcitation{AFT}{DMT00ApproximationsStableOperatorsWell-FoundedFixpointsApplications}

\setcitation{SAT}{BHvW21HandbookSatisfiability-SecondEdition}
\setcitation{HandbookOfSAT}{BHvW21HandbookSatisfiability-SecondEdition}
\setcitation{SMS}{KS24SATModuloSymmetriesGraphGenerationEnumeration}

\setcitation{totalizer}{BB03EfficientCNFEncodingBooleanCardinalityConstraints}

\setcitation{satsuma}{ABR26SatsumaStructure-BasedSymmetryBreakingSAT}
\setcitation{BreakID}{DBBD16ImprovedStaticSymmetryBreakingSAT}
\setcitation{symchaff}{S09SymChaffexploitingsymmetrystructure-awaresatisfiabilitysolver}

\setcitation{saucy}{KSM10SymmetrySatisfiabilityUpdate}
\setcitation{dejavu}{AS21EngineeringFastProbabilisticIsomorphismTest}
\setcitation{nauty}{MP14PracticalgraphisomorphismII}
\setcitation{bliss}{JK07EngineeringEfficientCanonicalLabelingToolLarge}

\setcitation{DPLLT}{GHNOT04DPLLTFastDecisionProcedures}
\setcitation{SMT}{BSST21SatisfiabilityModuloTheories}

\setcitation{CP}{RvW06HandbookConstraintProgramming}
\setcitation{csp2asp}{DW11Translation-BasedConstraintAnswerSetSolving}
\setcitation{EZCSP}{B11Industrial-SizeSchedulingASPCP}
\setcitation{Inca}{DW12AnswerSetSolvingLazyNogoodGeneration}
\setcitation{Zinc}{MNRSGW08DesignZincModellingLanguage}
\setcitation{MiniZinc}{NSBBDT07MiniZincTowardsStandardCPModellingLanguage}
\setcitation{Conjure}{AFGJMN22ConjureAutomaticGenerationConstraintModelsfrom}
\setcitation{Essence}{FHJHM08Essenceconstraintlanguagespecifyingcombinatorialproblems}

\setcitation{ASPComp2}{DVBGT09SecondAnswerSetProgrammingCompetition}
\setcitation{ASPComp3}{CIR14thirdopenanswersetprogrammingcompetition}
\setcitation{ASPComp4}{ACCDDIKK13FourthAnswerSetProgrammingCompetitionPreliminary}
\setcitation{ASPComp5}{CGMR16DesignresultsFifthAnswerSetProgramming}
\setcitation{ASPComp6}{GMR17SixthAnswerSetProgrammingCompetition}
\setcitation{ASPComp7}{GMR20SeventhAnswerSetProgrammingCompetitionDesign}
\setcitation{AspInPractice}{GKKS12AnswerSetSolvingPractice}
\setcitation{clasp}{GKS12Conflict-drivenanswersetsolvingFromtheory}
\setcitation{oclingo}{GGKOSS12StreamReasoningAnswerSetProgrammingPreliminary}
\setcitation{clingo}{GKKOSW16TheorySolvingMadeEasyClingo5}
\setcitation{gringo}{GST07GrinGoNewGrounderAnswerSetProgramming}
\setcitation{cmodels}{GLM04SAT-BasedAnswerSetProgramming}
\setcitation{DLV}{LPFEGPS06DLVsystemknowledgerepresentationreasoning}
\setcitation{GroundingBottleneck}{SPL14TwoApplicationsASP-PrologSystemDecomposablePrograms}
\setcitation{lazygroundingASP}{BW18ExploitingJustificationsLazyGroundingAnswerSet} 
\setcitation{justificationsAlpha}{BW18ExploitingJustificationsLazyGroundingAnswerSet}
\setcitation{ASP}{MT99StableModelsAlternativeLogicProgrammingParadigm}
\setcitation{OLL}{AKMS12Unsatisfiability-basedoptimizationclasp} 

\setcitation{Hilog}{CKW93HILOGFoundationHigher-OrderLogicProgramming}

\setcitation{MaxSAT}{LM21MaxSATHardSoftConstraints}

\setcitation{KR}{vLP08HandbookKnowledgeRepresentation}

\setcitation{lazyclausegeneration}{OSC09Propagationlazyclausegeneration}
\setcitation{FP}{H89ConceptionEvolutionApplicationFunctionalProgrammingLanguages}
\setcitation{GroundingWithBounds}{WMD10GroundingFOFOIDBounds}
\setcitation{GroundWithBounds}{WMD10GroundingFOFOIDBounds}
\setcitation{LTC}{BJBDVD14SimulatingDynamicSystemsUsingLinearTime}
\setcitation{SPSAT}{DBDDM12SymmetryPropagationImprovedDynamicSymmetryBreaking}
\setcitation{LCG}{S10LazyClauseGenerationCombiningPowerSAT}

\setcitation{GroundedFixpoints}{BVD15Groundedfixpointstheirapplicationsknowledgerepresentation}
\setcitation{PartialGroundedFixpoints}{BVD15PartialGroundedFixpoints}
\setcitation{LogicBlox}{GAK12LogicBloxPlatformLanguageTutorial}
\setcitation{proB}{LB08ProBautomatedanalysistoolsetBmethod}
\setcitation{NaturalInductions}{DV14Well-FoundedSemanticsPrincipleInductiveDefinitionRevisited} 
\setcitation{LP}{vK76SemanticsPredicateLogicProgrammingLanguage}

\setcitation{AF}{D95AcceptabilityArgumentsFundamentalRoleNonmonotonicReasoning}
\setcitation{ADF}{BW10AbstractDialecticalFrameworks}
\setcitation{ADFRevisited}{BSEWW13AbstractDialecticalFrameworksRevisited}
\setcitation{DefaultLogic}{R80LogicDefaultReasoning}
\setcitation{DL}{R80LogicDefaultReasoning}
\setcitation{AEL}{M85SemanticalConsiderationsNonmonotonicLogic}
\setcitation{minisat}{ES03ExtensibleSAT-solver}
\setcitation{completion}{C77NegationFailure}
\setcitation{ClarkCompletion}{C77NegationFailure}
\setcitation{wasp}{ADFLR13WASPNativeASPSolverBasedConstraint}
\setcitation{CEGAR}{CGJLV03Counterexample-guidedabstractionrefinementsymbolicmodelchecking}
\setcitation{CuttingPlane}{DFJ54SolutionLarge-ScaleTraveling-SalesmanProblem}
\setcitation{CuttingPlanes}{DFJ54SolutionLarge-ScaleTraveling-SalesmanProblem}
\setcitation{kodkod}{TJ07KodkodRelationalModelFinder}
\setcitation{CDCL}{SS99GRASPSearchAlgorithmPropositionalSatisfiability}
\setcitation{1UIP}{ZMMM01EfficientConflictDrivenLearningBooleanSatisfiability}
\setcitation{relevance}{JBDJD16RelevanceSATID}
\setcitation{relevance-implementation}{JDBJD16ImplementingRelevanceTrackerModule}
\setcitation{WFS}{VRS91Well-FoundedSemanticsGeneralLogicPrograms}
\setcitation{UnfoundedSet}{VRS91Well-FoundedSemanticsGeneralLogicPrograms}
\setcitation{UFS}{VRS91Well-FoundedSemanticsGeneralLogicPrograms}
\setcitation{stablesemantics}{GL88StableModelSemanticsLogicProgramming}
\setcitation{shatter}{ASM06EfficientSymmetryBreakingBooleanSatisfiability}
\setcitation{sbass}{DTW11Symmetry-breakinganswersetsolving}
\setcitation{lparsemanual}{url:lparsemanual}
\setcitation{AIC}{FGZ04Activeintegrityconstraints}
\setcitation{templates}{DvJD15Semanticstemplatescompositionalframeworkbuildinglogics}
\setcitation{templates2}{DvBJD16CompositionalTypedHigher-OrderLogicDefinitions}
\setcitation{sat-to-sat}{JTT16SAT-to-SATDeclarativeExtensionSATSolversNew}
\setcitation{sat-to-sat-qbf}{BJT16SolvingQBFInstancesNestedSATSolvers}
\setcitation{sat-to-sat-SO}{BJT16DeclarativeSolverDevelopmentCaseStudies}
\setcitation{XSB}{SW12XSBExtendingPrologTabledLogicProgramming}
\setcitation{KCmap}{DM02KnowledgeCompilationMap}
\setcitation{KnowledgeCompilationMap}{DM02KnowledgeCompilationMap}
\setcitation{TLA}{L02SpecifyingSystemsTLALanguageToolsHardware}
\setcitation{EventB}{A10ModelingEvent-B-SystemSoftwareEngineering}
\setcitation{MX}{MT05FrameworkRepresentingSolvingNPSearchProblems}
\setcitation{MIP}{S96Linearintegerprogramming-theorypractice}
\setcitation{PerfectModel}{P88DeclarativeSemanticsDeductiveDatabasesLogicPrograms}
\setcitation{SafeInductions}{BVD17SafeInductionsAlgebraicStudy}
\setcitation{alpha}{W17BlendingLazy-GroundingCDNLSearchAnswer-SetSolving}
\setcitation{omiga}{DEFWW12OMiGAOpenMindedGroundingOn-The-FlyAnswer}
\setcitation{gasp}{PDPR09GASPAnswerSetProgrammingLazyGrounding}
\setcitation{asperix}{LN09FirstVersionNewASPSolverASPeRiX}
\setcitation{CTL}{CE81DesignSynthesisSynchronizationSkeletonsUsingBranching-Time}
\setcitation{AFT-AIC}{BC18Fixpointsemanticsactiveintegrityconstraints}
\setcitation{UltimateApproximator}{DMT04Ultimateapproximationapplicationnonmonotonicknowledgerepresentation}
\setcitation{KripkeKleene}{F85Kripke-KleeneSemanticsLogicPrograms}
\setcitation{AFT-HO}{CRS18ApproximationFixpointTheoryWell-FoundedSemanticsHigher-Order} 
\setcitation{HereThere}{H30formalenRegelnintuitionistischenLogik}
\setcitation{dAEL}{VCBD16DistributedAutoepistemicLogicApplicationAccessControl}
\setcitation{SDD}{D11SDDNewCanonicalRepresentationPropositionalKnowledge}
\setcitation{HEX}{EIST06dlvhexProverSemantic-WebReasoningunderAnswer-Set}
\setcitation{wADF}{BSWW18WeightedAbstractDialecticalFrameworks}
\setcitation{wADFfix}{BPSWW18WeightedAbstractDialecticalFrameworksExtendedRevised}
\setcitation{TransitionSystems}{NOT06SolvingSATSATModuloTheoriesFrom}
\setcitation{galliwasp}{MG12GalliwaspGoal-DirectedAnswerSetSolver}
\setcitation{clingcon}{BKOS17Clingconnextgeneration}
\setcitation{lp2sat}{JN11CompactTranslationsNon-disjunctiveAnswerSetPrograms}
\setcitation{lp2mip}{LJN12AnswerSetProgrammingMixedIntegerProgramming}
\setcitation{lp2diff}{JNS09ComputingStableModelsReductionsDifferenceLogic}
\setcitation{lp2acyc}{GJR14AnswerSetProgrammingSATmoduloAcyclicity}
\setcitation{PB}{RM09Pseudo-BooleanCardinalityConstraints}
\setcitation{CuttingPlanes}{CCT87complexitycutting-planeproofs}
\setcitation{RoundingSAT}{EN18DivideConquerTowardsFasterPseudo-BooleanSolving}
\setcitation{PRS}{DG02InferenceMethodsPseudo-BooleanSatisfiabilitySolver}
\setcitation{sat4j}{LP10Sat4jlibraryrelease22}
\setcitation{mingo}{LJN12AnswerSetProgrammingMixedIntegerProgramming}
\setcitation{pbmodels}{LT05Pbmodels-SoftwareComputeStableModels}
\setcitation{HEF-LP}{GLL07Head-Elementary-Set-FreeLogicPrograms}
\setcitation{HCF-LP}{BD94PropositionalSemanticsDisjunctiveLogicPrograms}
\setcitation{aspcore2}{CFGIKKLM20ASP-Core-2InputLanguageFormat}
\setcitation{SemanticWeb}{AGvH12SemanticWebPrimer3rdEdition}
\setcitation{QBF}{KB21TheoryQuantifiedBooleanFormulas}
\setcitation{SMUS}{IPLM15SmallestMUSExtractionMinimalHittingSet}
\setcitation{CDCLsym}{MBCK18CDCLSymIntroducingEffectiveSymmetryBreakingSAT}
\setcitation{SEL}{DBB17SymmetricExplanationLearningEffectiveDynamicSymmetry}
\setcitation{Coq}{BC04InteractiveTheoremProvingProgramDevelopment-}
\setcitation{Isabelle}{NPW02IsabelleHOL-ProofAssistantHigher-OrderLogic}
\setcitation{HOL}{SN08BriefOverviewHOL4}
\setcitation{lean}{dU21Lean4TheoremProverProgrammingLanguage}
\setcitation{CakeML}{TMKFON19verifiedCakeMLcompilerbackend}
\setcitation{FRAT}{BCH22FlexibleProofFormatSATSolver-ElaboratorCommunication}
\setcitation{DRUP}{GN03VerificationProofsUnsatisfiabilityCNFFormulas}
\setcitation{RUP}{GN03VerificationProofsUnsatisfiabilityCNFFormulas}
\setcitation{GRIT}{CMS17EfficientCertifiedResolutionProofChecking}
\setcitation{TraceCheck}{url:TraceCheck}
\setcitation{LRAT}{CHHKS17EfficientCertifiedRATVerification}

\setcitation{PR}{HKB17ShortProofsWithoutNewVariables}
\setcitation{DRAT}{WHH14DRAT-trimEfficientCheckingTrimmingUsingExpressive}
\setcitation{satcomp2016}{BHJ17SATCompetition2016RecentDevelopments}
\setcitation{satcomp2013}{url:SATcomp2013}

\setcitation{smodels}{SN01SmodelsSystem}
\setcitation{lparse}{S98ImplementationLocalGroundingLogicProgramsStable}
\setcitation{IDPZ3}{CVVD22IDP-Z3reasoningengineFO} 
\setcitation{IDP-Z3}{CVVD22IDP-Z3reasoningengineFO} 
\setcitation{Z3}{dB08Z3EfficientSMTSolver}

\setcitation{DMN}{DMN} 
\setcitation{Planning}{GNT04Automatedplanning-theorypractice}
\setcitation{AutomatedPlanning}{GNT04Automatedplanning-theorypractice}
\setcitation{RC2}{IMM19RC2EfficientMaxSATSolver}
\setcitation{enfragmo}{phd/Aavani14}
\setcitation{TPTP}{url:tptp}

\newcommand\ourgrounder{\toolname{GroundFOX}}
\newcommand\ourchecker{\toolname{CheckFOX}}
\newcommand\ourtoolkit{\toolname{CertiFOX}}
\newcommand\ourproofformat{\toolname{CertiFOX}}
\newcommand\ournormalform{grounding normal form\xspace} 

\newcommand\OurNormalForm{Grounding Normal Form\xspace}
\newcommand\ourNF{GNF\@\xspace}

\lstdefinelanguage{certifox}{
  sensitive=true,
  extendedchars=true,
  escapeinside={(*}{*)},
  basicstyle={\small\ttfamily\upshape},
  columns=fullflexible,  
  backgroundcolor=\color{black!2},
  frame=single,
  mathescape=false,
  morecomment=[l]{//},
  commentstyle=\color{gray},
  otherkeywords={@, ->, -},
  keywords=[1]{
    IQ,
    SPLITC,
    EPRED,
    SNOR,
    SNAND,
    STN,
    TRIVIAL,
    UNSAT
  },
  keywords=[2]{
    FINAL,
    IDS
  },
  keywordstyle=[1]\color{blue!80!black},
  keywordstyle=[2]\color{orange!80!black},
}
\newcommand{\certifoxinline}[1]{{\lstinline[language=certifox]{#1}}}
\makeatletter
\lst@CCPutMacro
  \lst@ProcessOther {"7E}{$\sim$}
\@empty\z@\@empty
\makeatother

\title{Towards a Certifying Grounder\thanks{\thanksCertiFOX}}
\author{Daimy Van Caudenberg\institute{KU Leuven, Leuven, Belgium}\email{daimy.vancaudenberg@kuleuven.be}
\and Alexander Ek\institute{KU Leuven, Leuven, Belgium \\ ARC Training Centre OPTIMA, Melbourne, Australia}\email{alexander.ek@kuleuven.be}
\and Carlos Cantero\institute{KU Leuven, Leuven, Belgium}\email{carlos.cantero@kuleuven.be}
\and Bart Bogaerts\institute{KU Leuven, Leuven, Belgium \\ Vrije Universiteit Brussel, Brussels, Belgium}\email{bart.bogaerts@kuleuven.be }}

\newcommand{\titlerunning}{Towards a Certifying Grounder}
\newcommand{\authorrunning}{D. Van Caudenberg, A. Ek, C. Cantero, and B. Bogaerts}
\newcommand{\copyrightholders}{Van Caudenberg, Ek, Cantero, and Bogaerts}

\hypersetup{
  bookmarksnumbered,
  pdftitle    = {\titlerunning},
  pdfauthor   = {\authorrunning},
  pdfsubject  = {EPTCS},               
}

\begin{document}
\maketitle


\begin{abstract}
Grounding, the translation of high-level theories into equivalent quantifier-free formulas, is a crucial step in declarative solving, yet it has so far escaped the proof-logging revolution.
When this grounding step is not certifying, there is no way of knowing that the obtained solutions actually correspond to the original problem specification, resulting in a trust gap.
In this paper, we close the trust gap between the user's high-level specification and the solver's low-level input by introducing a novel certifying grounding framework for first-order logic model expansion (FOX) over finite domains.
We present CertiFOX, a framework consisting of: (1) a proof format for grounding derivations, (2) GroundFOX, a certifying grounder operating on theories in Grounding Normal Form (GNF)---a new normal form designed for compact, domain-aware grounding---and (3) CheckFOX, an independent proof checker.
Our approach guarantees that the grounder's output is equivalent to the input specification, setting the stage for trustworthy end-to-end certified solving pipelines for declarative languages.
Experimental evaluation confirms that \ourtoolkit is a feasible approach.
The \ourgrounder grounder is broadly comparable with other grounders, and proof checking with \ourchecker adds overhead within a small constant factor of grounding time.

\end{abstract}

\section{Introduction}
The field of combinatorial search and optimization is concerned with solving problems that are often NP-hard. In several subfields, this is done using declarative specifications in a suitable formal language.
Over the past decades, we have witnessed impressive improvements in the richness of the languages, the efficiency of solvers, and their industrial applications.
Since some of these applications involve high-value and life-affecting decision-making processes (e.g., validating the correctness of plans for space shuttle operation~\cite{NBGWB01A-PrologDecisionSupportSystemSpaceShuttle},
or matching donors and recipients for kidney transplants~\cite{MO14PairedAltruisticKidneyDonationUKAlgorithms}), it is of utmost importance that the answers produced by the solvers be 
reliable.

Unfortunately, the reality is different: the constant need for more efficient and advanced algorithms creates an excellent breeding ground for bugs, resulting in numerous reports of bugs in solvers and of solvers outputting faulty answers~\cite{BB09FuzzingDelta-DebuggingSMTSolvers,BLB10AutomatedTestingDebuggingSATQBFSolvers,JHB12InprocessingRules,CKSW13hybridbranch-and-boundapproachexactrationalmixed-integer,AGJMN18MetamorphicTestingConstraintSolvers,GSD19SolverCheckDeclarativeTestingConstraints}.
The question that naturally arises is how to get stronger guarantees of correctness of a solver.
More specifically, if a solver claims that a problem has no solutions, how can we know this is indeed the case? Or if a solver claims a specific solution is optimal, how can we be sure that there are no better solutions?

One successful way to achieve such a guarantee is the use of \emph{certifying algorithms}~\cite{ABMRS11IntroductionCertifyingAlgorithms,MMNS11Certifyingalgorithms}.
In the context of combinatorial search, this is also known as proof logging.
The idea is that algorithms (in our case for solving decision or optimization problems) should not just output an answer, but also a \emph{certificate} (or \emph{proof}) that this answer is correct.
The produced certificate should be sufficiently simple to be verified efficiently by an independent tool, often referred to as the \textbf{proof checker}.
This checker is a tool of much lesser complexity than the solver. Indeed, the checker does not need any complicated heuristics or search strategies; its sole job is confirming correctness of the derivations made by the solver.

Proof logging has been embraced strongly by the Boolean satisfiability (SAT) community; a wide variety of proof formats have been proposed, the most notable and commonly used one being DRAT~\cite{HHW13Trimmingwhilecheckingclausalproofs}, and proof logging has been mandatory for solvers participating in the annual SAT competition for years.
Inspired by this, proof logging is now seeing adoption in other fields of combinatorial optimization, such as satisfiability modulo theories (SMT)~\cite{BRKLNNOP22FlexibleProofProductionIndustrial-StrengthSMTSolver}, (standard and multi-objective) MaxSAT~\cite{VCB26CertifiedBranch-and-BoundMaxSATSolving,JBBJ25CertifyingParetoOptimalityMulti-ObjectiveMaximumSatisfiability}, constraint programming~\cite{GMN22AuditableConstraintProgrammingSolver,FSMSD24Multi-StageProofLoggingFrameworkCertifyCorrectness}, classical planning~\cite{ERH17UnsolvabilityCertificatesClassicalPlanning} and answer set programming~\cite{ADFHPR19InconsistencyProofsASPASP-DRUPE}, thereby supporting advanced solving techniques~\cite{BGMN23CertifiedDominanceSymmetryBreakingCombinatorialOptimisation,IVSBBJ26EfficientReliableHitting-SetComputationsImplicitHitting}.
Proofs of this form are not only useful for guaranteeing the correctness of the solver's answer, but also for debugging and testing solvers~\cite{BBNOV23CertifiedCore-GuidedMaxSATSolving}, and for explainability purposes~\cite{BFBDG26UsingCertifyingConstraintSolversGeneratingStep-wise}.

In several domains, including constraint programming and answer set programming, users often specify problems in a high-level, human-readable language, which is then translated (by a technique called \emph{grounding}) into a low-level format that a solver can parse.
One main limitation of proof logging research so far is that it has focused solely on the solving phase, meaning that the produced certificate only guarantees that the solver's answer is correct with respect to the input it receives, but not that the input itself is correct with respect to the original problem specification, resulting in a \emph{trust gap}.

The goal of this paper is to bridge this gap.
Concretely, we approach this problem from a logical perspective and focus on the \emph{model expansion}~\mycite{MX} problems for first-order logic (FOL); in brief, we will write FOX below to refer to first-order logic model expansion.
A model expansion problem takes as input a theory in first-order logic, representing domain knowledge, and a structure interpreting a subset of the symbols, representing the concrete instance of the problem at hand.
The goal is to expand this input structure to a model of the theory, which represents a solution to the problem.

Typically, a FOX problem is solved in two phases: first, the input theory is \emph{grounded} to an equivalent quantifier-free theory, and then this grounded theory is solved using a suitable solver (e.g., a SAT solver).
In practice, state-of-the-art grounders~\cite{WMD10GroundingFOFOIDBounds,\refto{lparse},\refto{gringo}} apply sophisticated optimizations and rewrite rules to handle large domains efficiently, making them highly complex and difficult to verify.

\paragraph{Contributions}
In this paper, we present the foundations of the \ourtoolkit ecosystem, a certifying framework for the grounding of FOX specifications.
The \ourtoolkit ecosystem consists of three main components: the \ourproofformat proof format, the \ourgrounder certifying grounder, and the \ourchecker proof checker.
We focus on a fragment of first-order logic in which sentences are expressed in \emph{\ournormalform} (\ourNF), a new normal form designed to exploit domain knowledge to produce compact groundings.
The goal of the \ourtoolkit ecosystem is to provide a methodology for the certifiable grounding of FOX specifications in \ourNF, closing the trust gap between the user's specification and the solver's input.

\paragraph{Related Work}
Proof logging has been studied extensively as an approach to certifying solvers, resulting in many proof formats such as DRAT for SAT~\cite{HHW13Trimmingwhilecheckingclausalproofs}; VeriPB for pseudo-Boolean solving~\cite{KLMNOTV25PracticallyFeasibleProofLoggingPseudo-BooleanOptimization,IVSBBJ26EfficientReliableHitting-SetComputationsImplicitHitting}, MaxSAT~\cite{VDB22QMaxSATpbCertifiedMaxSATSolver,BBNOV23CertifiedCore-GuidedMaxSATSolving}, and constraint programming~\cite{GMN22AuditableConstraintProgrammingSolver}; and eDRAT for SMT solving~\cite{Hitarth2024}.
By contrast, the grounding phase has received little attention from a verification standpoint.
The certification of preprocessing steps such as translations is the closest analogue to our work: preprocessing rewrites a formula before solving, much as grounding rewrites a high-level theory before solving.
Certifying preprocessing steps has been studied for example in the context of
verified translations~\cite{GMNO22CertifiedCNFTranslationsPseudo-BooleanSolving}, MaxSAT~\cite{IOTBJMN24CertifiedMaxSATPreprocessing}, and quantified Boolean formulas~\cite{HSB14UnifiedProofSystemQBFPreprocessing}.
SMT is particularly relevant, because its preprocessing also encompasses quantifier instantiation, and dedicated proof rules exist to certify these instantiation steps~\cite{DeharbeFP11,BBFF20ScalableFine-GrainedProofsFormulaProcessing,NBNPRT22ReconstructingFine-GrainedProofsRewritesDSL}.
Unlike grounding, however, SMT instantiation is non-exhaustive---it produces only enough ground instances to satisfy the solver, rather than an equivalent quantifier-free problem.

Our work extends these certifying approaches to include the grounding of first-order logic,
closing the trust gap between the user's specification and the solver's input.
This is particularly important in the context of high-level modeling languages, where users rely on the grounder to faithfully translate their specifications into a form suitable for solving.
\idpz~\cite{\refto{IDPZ3}} and \textsc{clingo}~\cite{\refto{clingo}} are high-level modeling
systems that perform grounding for \fodot and ASP respectively; neither currently produces certificates of correctness for their translations, meaning the final solutions can only be trusted insofar as the grounding process can be trusted.
Because both systems are highly complex software systems with many optimizations and rewrite rules, it is difficult to fully verify their correctness, and thus there is a risk of bugs leading to incorrect groundings and, by extension, incorrect final solutions.

\section{Preliminaries}
A combinatorial search problem asks whether there is an object, within a finite domain, that satisfies all given constraints.
For example, given a CNF formula, the problem of finding a satisfying assignment is a combinatorial search problem.

\paragraph{First-Order Logic}
These preliminaries are based on~\cite{WMD10GroundingFOFOIDBounds}.
A \emph{vocabulary} $\voc$ is a set of predicate and function symbols. Each predicate and function symbol has an \emph{arity}, which is the number of arguments it takes.
We denote predicate (function) symbols with arity $n$ using $\fn{P}{n}$ ($\fn{f}{n}:$).
A predicate (function) with arity $0$ is called a \emph{proposition} (\emph{object}) symbol.
We assume that every vocabulary contains the propositions $\ltrue$ and $\lfalse$, representing truth and falsity, respectively.

A \emph{term} is a variable or an $n$-ary function applied to $n$ terms.
An \emph{atom} is an $n$-ary predicate applied to $n$ terms.
If $p$ and $q$ are terms, $p=q$ is also an atom.
Next, we define the \emph{formulas}:
\begin{itemize}
	\item an atom is a formula,
	\item if $\phi$ is a formula, then so is $\lnot\phi$,
	\item if $\phi$ and $\psi$ are formulas, then so are $\phi\imply\psi$, $\phi\limply\psi$ and $\phi\ifff\psi$,
	\item if $\phi_i$ is a formula for all $i\in\{1,\ldots,n\}$, then so are $\bigwedge_{i=1}^n \phi_i$ and $\bigvee_{i=1}^n \phi_i$,
	\item if $\phi$ is a formula and $x$ is a variable then $\forall x:\phi$ and $\exists x:\phi$ are formulas.
\end{itemize}
A \emph{sentence} is a formula without free variables.
A \emph{theory} $\theory$ over vocabulary $\voc$ is a set of sentences over the symbols of $\voc$.
A \emph{structure} $\struct$ over vocabulary $\voc$ consists of a domain $\domain$ and an interpretation for all symbols in $\voc$, where for $n$-ary predicate symbols, the interpretation is a subset of $\domain^n$, for $n$-ary function symbols, the interpretation is a function from $\domain^n$ to $\domain$.
We assume that the domain $\domain$ is finite.
A structure $\struct$ over  $\voc$ \emph{expands} a structure $\struct'$ over $\voc'\subseteq\voc$ if they have the same domain and $\struct$ agrees with $\struct'$ on all symbols in $\voc'$ interpreted by $\struct'$; this is denoted using $\struct \geqp \struct'$ (here the $\mathrm{p}$ indicates that the first structure is more precise than the second).
A structure $\struct$ \emph{satisfies} a sentence $\phi$ if $\struct$ makes $\phi$ true under the standard semantics of FOL\@; this is denoted as $\struct\models\phi$.
A structure $\struct$ is a \emph{model} of a theory $\theory$ if it satisfies all sentences in $\theory$.

\paragraph{Model Expansion}
A model expansion problem is a combinatorial search problem where a structure over a subset of the vocabulary is given (that is, the \emph{input vocabulary}), and the goal is to find a structure that expands it to the full vocabulary (which includes the \emph{output vocabulary}) and satisfies a given theory.
More formally, a First-Order Logic model expansion problem is a tuple $\FOXProbshort$ consisting of:
  \begin{itemize}
    \item a vocabulary $\voc=\inputVocabulary\cup\outputVocabulary$ which is the union of the input and output vocabularies,
    \item a structure $\inputStructure$ over $\inputVocabulary$ providing interpretations for input symbols,
    \item a theory $\theory$ over $\voc$.
  \end{itemize}
The \emph{FOL model expansion problem} (FOX) consists of finding a structure $\struct$ such that:
\begin{itemize}
	\item $\struct\geqp\inputStructure$ (meaning $\struct$ expands $\inputStructure$),
	\item and $\struct\models\theory$ (meaning it is a model of $\theory$).
\end{itemize}
We illustrate the flexibility of FOX model specifications with two toy examples.
\begin{example}\label{ex:ph}
    We now define the pigeonhole problem as a model expansion problem.
    First, we define the vocabulary $\voc=\{\pigeon/1,\hole/1,\sit/2\}$, which contains predicates that allow us to distinguish pigeons from holes, and to assign pigeons to holes.
    Next, we define the theory:
    \begin{align*}
        \theory=
        \begin{Bmatrix}
            \forall p: \pigeon(p) \imply \exists h: \hole(h)\land\sit(p,h).\\
            \forall p_1, p_2, h: \left(\hole(h) \land \pigeon(p_1) \land \pigeon(p_2) \land p_1\neq p_2\right) \imply \lnot\sit(p_1,h) \lor\lnot\sit(p_2,h).
        \end{Bmatrix}
    \end{align*}
    The sentences ensure that each pigeon sits in a hole and that each hole can contain at most one pigeon.
    Given the structure $\struct=\{\domain=\{P_1,P_2,H\},\pigeon=\{P_1,P_2\},\hole=\{H\}\}$ with fewer holes than pigeons, model expansion of the structure $\struct$ given theory $\theory$ fails.
    Note that for $\struct'=\{\domain=\{P_1,P_2,H\},\pigeon=\{P_1,P_2\},\hole=\{P_1,H\}\}$, model expansion succeeds because the theory does not enforce correct typing.
\end{example}
\begin{example}\label{ex:connected}
    The goal is to define a model expansion problem that finds a graph containing a node connected to all other nodes.
    First, we define the vocabulary $\voc=\inputVocabulary\cup\outputVocabulary$, where the input vocabulary $\inputVocabulary=\{\Node/1\}$ contains a predicate that represents nodes, and the output vocabulary contains a predicate for edges ($\outputVocabulary=\{\Edge/2\}$).
    Next, we define the theory $\theory$ as follows:
    \begin{align*}
      \theory=
      \begin{Bmatrix}
        \exists x: \Node(x) \land \forall y: \Node(y) \land y\neq x \imply \Edge(x,y).\\
      \end{Bmatrix}
    \end{align*}
    The sentence ensures that there exists a node $x$ such that for all other nodes $y$, there is an edge from $x$ to $y$.
    Model expansion of the structure $\struct=\{\domain= 
    \Node=\{N_1,N_2,N_3,N_4\}\}$ and  theory $\theory$
    will succeed.
    For instance,  $\struct'=\{\domain= 
    \Node=\{N_1,N_2,N_3,N_4\}$, $\Edge=\{(N_1,N_2),(N_1,N_3),(N_1,N_4),(N_2,N_3)\}\}$ is a structure that satisfies the theory; indeed, it contains the node $N_1$, which is connected to all other nodes.
  \end{example}

\section{\ourtoolkit Ecosystem}
\begin{figure}
  \centering
  \resizebox{0.4\textwidth}{!}{
    \begin{tikzpicture}[
    every node/.style={font=\small},
    tool/.style={draw, rounded corners=3mm, minimum width=2.5cm, align=center,inner sep=5pt},
    proof/.style={draw,  minimum width=2.5cm, align=center,inner sep=5pt},
    arrow/.style={->, very thick},
    node distance=0.5cm,
  ]
                    \node[tool] (ground) {\ourgrounder\\Grounder};

                    \node[proof, below= of ground] (proofground) {\ourtoolkit\\Proof};
                    \node[tool, below= of proofground] (checkground) {\ourchecker\\Proof Checker};
                    \node[draw=none,below=of checkground] (checksground) {\Large\textcolor{lightgreen}{\ding{51}}/\textcolor{red}{\ding{55}}};

                   \draw[arrow] (ground) -- ++(3.5cm,0pt) node[above,midway,align=center]{Low-level\\specification\\(CNF)};

                   \draw[<-, very thick] (ground) -- ++(-3.5cm,0pt) node[above,midway,align=center]{High-level\\specification\\(FOX)};

                   \draw[<-, very thick] (checkground) -| ($(ground)+(2cm,0pt)$);
                   \draw[<-, very thick] (checkground) -| ($(ground)+(-2cm,0pt)$);
                  \draw[arrow] (ground) -- (proofground);
                  \draw[arrow] (proofground) -- (checkground);
                  \draw[arrow] (checkground) -- (checksground);

               \end{tikzpicture}
  }
               \caption{Architecture of the prototype.}
               \label{fig:arch}

\end{figure}
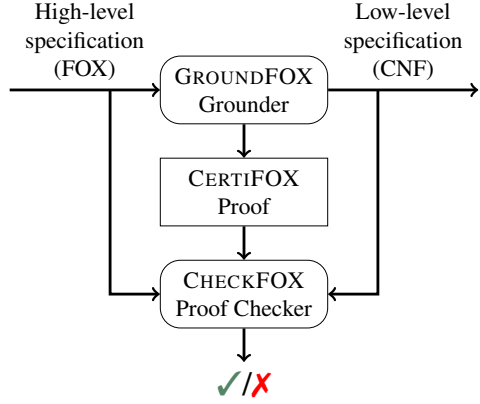

The \ourtoolkit ecosystem is a novel methodology for the certifiable grounding of FOX specifications.
The \textbf{\ourgrounder grounder} takes as input a FOX problem (in a specific normal form defined below), and grounds it to an equivalent CNF formula.
During the grounding process, each transformation step is logged in a \textbf{\ourproofformat proof}, producing a proof certificate that can be independently verified.
The checker then uses the original FOX formula, the grounded CNF formula, and the proof to validate the correctness of the grounding process.
This separation of concerns ensures that bugs cannot compromise soundness, as the checker provides an independent verification layer.
The architecture is illustrated in \cref{fig:arch}, showing the data flow from input formula through grounding and certification.

\subsection{\OurNormalForm}
Our prototype of the \ourtoolkit ecosystem is limited to a fragment of FOL in which sentences are expressed in \emph{grounding normal form} (GNF).
GNF is a normal form that allows grounding to conjunctive normal form (CNF) after quantifier instantiation and basic simplifications (without the need for the introduction of auxiliary symbols by the grounder).
By grounding directly to CNF, users can leverage the power of certifying SAT solvers without needing an additional translation step.
This restriction allows us to focus our efforts on certifying the grounding process itself, as opposed to the translation from the grounded formula to conjunctive normal form.

Our normal form is designed to allow for compact grounding by taking advantage of available domain knowledge.
This is done by introducing \emph{binary} quantifiers (inspired by~\cite{Michael1995}), which allow users to define \emph{guards} for the quantifiers.
These quantifiers are called \emph{binary} because they take two formulas as arguments: the first is the guard, and the second is the formula being quantified over.
\begin{definition}[Binary Quantifiers]
    \label{def:binquant}
    Given formulas $\varphi$ and $\psi$, the binary quantifiers used in GNF are:
    \begin{align*}
        \forall x [\varphi(x)] : \psi(x) \define \forall x : \varphi(x) \Rightarrow \psi(x) \\
        \exists x [\varphi(x)] : \psi(x) \define \exists x : \varphi(x) \land \psi(x).
    \end{align*}
    We call $\varphi$ a \textit{guard} for $\psi$ and write $Q x:\psi(x)$ as an abbreviation for $Q x[\ltrue]:\psi(x)$, where $Q\in\{\forall, \exists\}$ and $\ltrue$ is the symbol for truth.
\end{definition}
Given a FOX problem $\FOXProb$, it is required that guards of the quantifiers in $\theory$ only contain symbols from $\inputVocabulary$, which are interpreted by the user in $\inputStructure$.
This allows the grounder to evaluate the guards during the grounding phase, and to skip irrelevant ground instances.

We now introduce the grounding normal form (GNF), which uses these binary quantifiers.
\begin{definition}[Grounding Disjunctive Form]
  \
  \begin{itemize}
    \item If $\psi$ is an atom or a negated atom, then $\psi$ is in grounding disjunctive form.
    \item If $\psi$ and $\phi$ are in grounding disjunctive form, then $(\psi \lor \phi)$ is in grounding disjunctive form.
    \item If $\psi$ is in grounding disjunctive form and $\phi$ is a FO-formula all symbols of which are in $\inputVocabulary$, then
        $(\exists x [\phi] : \psi)$ is in grounding disjunctive form.
  \end{itemize}
\end{definition}
\begin{definition}[Grounding Normal Form]
    \
    \begin{itemize}
        \item If $\psi$ is in grounding disjunctive form, then $\psi$ is in grounding normal form (GNF).
        \item If $\psi$ and $\phi$ are in GNF, then $(\phi \land  \psi)$ is in \ourNF.
        \item If $\psi$ is in GNF and $\varphi$ is a FO-formula all symbols of which are in $\inputVocabulary$, then $(\forall x [\varphi] : \psi)$ is in \ourNF.
    \end{itemize}
\end{definition}
Note that every FOL formula can be transformed into an equisatisfiable \ourNF formula.
When the quantifiers in the original formula are ordered in a way that is not compatible with \ourNF, this transformation may require the introduction of auxiliary symbols.
If this is not the case, as in \cref{ex:ph}, then the transformation requires only the definition of binary quantifiers and logical equivalences. We illustrate this translation process by transforming the theory from \cref{ex:connected} into \ourNF.
\begin{example}\label{ex:connected-gnf}
  We transform the theory from \cref{ex:connected} into \ourNF.
  The original sentence is not in \ourNF because it contains an existentially quantified universal quantifier.
  We introduce an auxiliary symbol $\Connected/1$, representing that a node is connected to all other nodes: \[\forall x: \Connected (x) \ifff \Bigl(\forall y: y\neq x \imply \Edge(x,y)\Bigr).\]
  We apply standard logical rewrites and use the definition of binary quantifiers to obtain the following equisatisfiable \ourNF theory:
    \begin{align*}
      \theory=
      \begin{Bmatrix}
          \exists x: \Connected(x).\\
          \forall x: \forall y\bigl[ y\neq x \bigr]: \Edge(x,y) \lor \lnot\Connected(x).\\
      \forall x: \exists y\bigl[ y\neq x \bigr]: \lnot\Edge(x,y) \lor \Connected(x).
      \end{Bmatrix}
    \end{align*}
  \end{example}

\subsection{\ourproofformat Proof Format}\label{sec:proof}
The goal of a \ourproofformat proof is to certify that a grounder correctly transformed the FOX problem $\FOXProbshort$ into an equivalent CNF formula $\theory'$, by providing a machine-checkable proof of the equivalence between $\theory$ and $\theory'$.
The proof format is built on a minimal collection of sound, equivalence-preserving rewrite rules, the most important being the instantiation rules for binary quantifiers.
We first discuss the rules of the proof system, and then we argue informally that each rule preserves equivalence (given the input structure), which is the key correctness guarantee of \ourtoolkit.
Next, we describe the syntax of the proof format, and we illustrate it with an example.

\paragraph{\ourproofformat Rules}
Because the grounder can exploit the known interpretation of input symbols provided by $\inputStructure$, the proof system is designed to preserve a weaker notion of equivalence, called $\inputStructure$-equivalence.
\begin{definition}[$\inputStructure$-equivalence]
  Given a FOX problem $\FOXProb$, two formulas $\varphi$ and $\psi$ are \textbf{$\inputStructure$-equivalent}, written $\varphi \equiv_{\inputStructure} \psi$ if for every structure $\struct$ extending $\inputStructure$, it holds that $\struct$ is a model of $\varphi$ if and only if it is a model of $\psi$ (i.e., $\forall \struct\geqp\inputStructure: \struct \models \varphi \ifff \struct \models \psi$).
  By extension, $\theory\equiv_{\inputStructure}\theory'$ means that the theories $\theory,\theory'$ are $\inputStructure$-equivalent, i.e., that they have the same models among the structures that extend $\inputStructure$.
\end{definition}

The correctness guarantee of the \ourtoolkit ecosystem is that given an input problem $\FOXProbshort$, every rule application preserves $\inputStructure$-equivalence at the theory level, meaning that if $\theory$ is the theory before applying the rule and $\theory'$ is the theory after applying the rule, then $\theory \equiv_{\inputStructure} \theory'$.
Because most of the rules supported by \ourproofformat are translations of well-known logical equivalences, we will not provide a formal proof of this guarantee for each rule. Instead, we provide an informal argument for each rule, which should be sufficient to convince the reader of the soundness of the proof system.

\cref{tab:formrules,tab:theoryrules} contain an overview of the rules of the \ourtoolkit proof system; \cref{tab:formrules} contains rules that apply at the level of a single (sub)formula, while \cref{tab:theoryrules} contains rules that apply at the level of the whole theory. 
\cref{tab:formrules} contains several rules that allow for the simplification of formulas, such as \snand, \snor, and \stn.
They are basic logical equivalences that hence also preserve $\inputStructure$-equivalence.
The rules \epred and \eprop allow for the interpretation of input predicates and propositions, respectively, by replacing them with $\ltrue$ or $\lfalse$ according to their interpretation in $\inputStructure$.
Because this interpretation is fixed, performing this replacement preserves all models among the structures that extend $\inputStructure$; hence it preserves $\inputStructure$-equivalence.
The last formula-level rules are the instantiation rules for binary quantifiers (\iq).
Concretely, given a formula of the form $Q x[\varphi(x)]: \psi(x)$ with $Q\in\{\forall,\exists\}$, the \iq-rule instantiates the formula only with values $v$ in the domain that satisfy the guard $\varphi$ according to the input structure  $\inputStructure$.
This is a crucial optimization, as it allows the grounder to skip irrelevant ground instances.
When taking the definition of binary quantifiers into account, it is not hard to see that this rule also preserves $\inputStructure$-equivalence.
\begin{table}[t]
  \centering
  \begin{NiceTabular}{cc}[
    cell-space-top-limit=3pt,
    cell-space-bottom-limit=3pt,
  ]
    $\vcenter{\hbox{%
      \AXC{$\bigwedge\limits_{i=1\dots n} \varphi_i$}
      \LL{\snand}
      \RL{$\varphi_i\neq \lfalse$ for all $i$}
      \UIC{$\bigwedge\limits_{\substack{i=1\dots n \\ \varphi_i \neq \ltrue}} \varphi_i$}
      \DisplayProof
    }}$
    &
    $\vcenter{\hbox{
      \AXC{$\bigwedge\limits_{i=1\dots n} \varphi_i$}
      \LL{\snand}
      \RL{$\varphi_i = \lfalse$ for some $i$}
      \UIC{$\vphantom{\bigwedge\limits_{\substack{i=1\dots n \\ \varphi_i \neq \ltrue}} \varphi_i}\lfalse$}
      \DisplayProof
    }}$
    \\
    $\vcenter{\hbox{
      \AXC{$\bigvee\limits_{i=1\dots n} \varphi_i$}
      \LL{\snor}
      \RL{$\varphi_i = \ltrue$ for some $i$}
      \UIC{$\vphantom{\bigvee\limits_{\substack{i=1\dots n \\ \varphi_i \neq \ltrue}} \varphi_i}\ltrue$}
      \DisplayProof
    }}$
    &
    $\vcenter{\hbox{%
      \AXC{$\bigvee\limits_{i=1\dots n} \varphi_i$}
      \LL{\snor}
      \RL{$\varphi_i\neq \ltrue$ for all $i$}
      \UIC{$\bigvee\limits_{\substack{i=1\dots n \\ \varphi_i \neq \lfalse}} \varphi_i$}
      \DisplayProof
    }}$
    \\
    $\vcenter{\hbox{%
      \AXC{$P(\bar{t})$}
      \AXC{$\inputStructure \models P(\bar{t})$}
      \LL{\textsc{epred}}
      \BIC{$\ltrue$}
      \DisplayProof
    }}$
    &
    $\vcenter{\hbox{%
      \AXC{$P(\bar{t})$}
      \AXC{$\inputStructure \not\models P(\bar{t})$}
      \LL{\textsc{epred}}
      \BIC{$\lfalse$}
      \DisplayProof
    }}$
    \\
    $\vcenter{\hbox{%
      \AXC{$P$}
      \AXC{$\inputStructure \models P$}
      \LL{\textsc{eprop}}
      \BIC{$\ltrue$}
      \DisplayProof
    }}$
    &
    $\vcenter{\hbox{%
      \AXC{$P$}
      \AXC{$\inputStructure \not\models P$}
      \LL{\textsc{eprop}}
      \BIC{$\lfalse$}
      \DisplayProof
    }}$
    \\
    $\vcenter{\hbox{%
      \AXC{$\forall x[\varphi(x)]: \psi(x)$}
      \LL{\textsc{IQ}}
      \UIC{$\bigwedge\limits_{\substack{v \in \domain\\\inputStructure \models \varphi(v)}} \psi(v)$}
      \DisplayProof
    }}$
    &
    $\vcenter{\hbox{%
      \AXC{$\exists x[\varphi(x)]: \psi(x)$}
      \LL{\textsc{IQ}}
      \UIC{$\bigvee\limits_{\substack{v \in \domain\\\inputStructure \models \varphi(v)}} \psi(v)$}
      \DisplayProof
    }}$
    \\
    $\vcenter{\hbox{%
      \AXC{$\lnot \ltrue$}
      \LL{\textsc{stn}}
      \UIC{$\lfalse$}
      \DisplayProof
    }}$
    &
    $\vcenter{\hbox{%
      \AXC{$\lnot \lfalse$}
      \LL{\textsc{stn}}
      \UIC{$\ltrue$}
      \DisplayProof
    }}$
    \\
  \end{NiceTabular}
  \caption{\ourtoolkit formula-level proof rules}
  \label{tab:formrules}
\end{table}

\cref{tab:theoryrules} presents the theory-level rules of the \ourtoolkit proof system.
Since a theory can be viewed as the conjunction of its formulas, the rules \textsc{TRIVIAL} and \textsc{UNSAT} provide theory-level counterparts of \textsc{SNAND}.
Concretely, a formula in the theory that reduces to $\ltrue$ may be removed (\textsc{TRIVIAL}), while one that reduces to $\lfalse$ allows early termination with unsatisfiability (\textsc{UNSAT}).
Note that the \unsat-rule can be used to show the unsatisfiability of the original problem, as it guarantees that there are no models among the structures that extend $\inputStructure$.
The \textsc{SPLITC} rule also operates at the theory level and decomposes a conjunctive formula into its conjuncts.
It is easy to see that all of these rules preserve $\inputStructure$-equivalence, as they are all based on well-known logical proof rules.

\begin{table}[t]
\centering
$\vcenter{\hbox{%
  \AXC{$\theory \cup \{\ltrue\}$}
  \LL{\textsc{trivial}}
  \UIC{$\theory$}
  \DisplayProof
}}$%
\qquad%
$\vcenter{\hbox{%
  \AXC{$\theory \cup \{\lfalse\}$}
  \LL{\textsc{unsat}}
  \UIC{$\textsc{abort}$}
  \DisplayProof
}}$%
\qquad%
$\vcenter{\hbox{%
  \AXC{$\theory \cup \{\varphi_1 \land \ldots\land \varphi_n\}$}
  \LL{\textsc{splitc}}
  \UIC{$\theory \cup \{\varphi_1,\ldots,\varphi_n\}$}
  \DisplayProof
}}$
\caption{\ourtoolkit theory-level proof rules}
\label{tab:theoryrules}
\end{table}

\paragraph{Positioning System}
Each rule application targets a specific (sub)formula using a custom positioning system built around the formula's syntax tree, ensuring that the proof is unambiguous but brief.
The positioning system is crucial for the proof format, as it allows the checker to precisely identify which (sub)formula to apply each rule to.
Concretely, a position is a tuple $\langle N,P\rangle$, consisting of a formula name $N$ and a sequence of indices $P=[i_0, i_1, \ldots, i_n]$.
The elements $i_j\in\mathbb{N}$ are called indices, and they identify a path from the root of that formula's syntax tree to the subformula to rewrite.
A tuple $\langle N,P\rangle$ can also be written as $N[i_0, i_1, \ldots, i_n]$; a tuple with an empty path is written as $N$, referring to the formula named $N$ itself.
Using \cref{tab:positionsystem}, we illustrate how this indexing system can be used to refer to specific subformulas and subterms.
In the second formula of \cref{ex:connected-gnf} (assuming it is named $2$), the position $\langle 2,[0]\rangle$ refers to the subformula $\forall \bigl[x\neq y\bigr]: \Edge(x,y) \lor \lnot\Connected(x)$.
The position $\langle 2,[0,0]\rangle$ refers to the guard of the quantifier inside that formula (i.e., $x\neq y$), while the position $\langle 2,[0,1]\rangle$ refers to the body of that quantifier (i.e., $\Edge(x,y) \lor \lnot\Connected(x)$).

\begin{table}[t]
\begin{center}
\begin{tabular}{l|l}
\textbf{Formula $\phi$} & \textbf{Positions} \\ \hline
$\neg \varphi$ & $\langle\phi,0\rangle=\varphi$ \\
$\varphi_1 \circ \varphi_2$ & $\langle\phi,0\rangle=\varphi_1$ (lhs), $\langle\phi,1\rangle=\varphi_2$ (rhs) \\
$\bigcirc(\varphi_0, \ldots, \varphi_n)$ & $\langle\phi,i\rangle=\varphi_i$ for $i \in [0,n]$ \\
$Qx[\psi]:\varphi$ & $\langle\phi,0\rangle=\psi$ (guard), $\langle\phi,1\rangle=\varphi$ (body) \\
$P(t_0, \ldots, t_n)$ & $\langle\phi,i\rangle=t_i$ for $i \in [0,n]$ \\
$t_1 = t_2$ & $\langle\phi,0\rangle=t_1$, $\langle\phi,1\rangle=t_2$ \\
$f(t_0, \ldots, t_n)$ & $\langle\phi,i\rangle=t_i$ for $i \in [0,n]$ \\
0-ary predicates and terms & No subformulas or subterms
\end{tabular}
\caption{The positioning system for referring to subformulas and subterms, where $\circ \in \{\Leftarrow, \Rightarrow, \Leftrightarrow \}$, $\bigcirc \in \{ \land, \lor \}$, and $Q \in \{ \forall, \exists \}$.}
\label{tab:positionsystem}
\end{center}
\end{table}

\paragraph{\ourproofformat Syntax}
A \ourproofformat proof consists of three main parts: a header containing version information, the body containing a chronological sequence of applied rewrite rules, and a footer containing identifiers for the obtained grounded formulas.
The rewrite rules allow the grounder to log its reasoning in a way that is easy to understand and verify.
With the positioning system in place, we can now describe the syntax of the \ourproofformat proof format.
The header specifies the version of the proof format and the grounder that produced the proof.

After the header, the body of the proof contains a chronological sequence of applied rewrite rules, which together form a complete record of the transformations performed by the grounder.
Each line in the proof corresponds to a single rule application, and is logged as in \cref{tab:proofsyntax}, where each rule application specifies the rule name, the position of the (sub)formula to target and for \splitc, the names of the new formulas added to the theory after splitting.
The footer of the proof contains identifiers for the obtained grounded formulas, which can be used by the checker to verify that the final formula is syntactically identical to the claimed grounding.
It is logged using \certifoxinline{FINAL IDS : <N>, ... <N>}, listing the names of the formulas that are present in the final grounded theory in the order they appear in the theory.
If the final grounded theory is empty, this line is logged as \certifoxinline{FINAL IDS : -}.
An illustration of a syntactically correct \ourproofformat proof is given in \cref{lst:proof}.

\begin{figure}[t]
\begin{lstlisting}[language=certifox,caption={The syntax of \ourproofformat rule applications.}, label={tab:proofsyntax}]
// Rules with @ modify the targeted (sub)formula - SNAND, SNOR, STN, EPRED, IQ, UNSAT
<rule name> @ <position>

// Rules with - delete the targeted formula - TRIVIAL
<rule name> - <position>

// Rules with -> replace the targeted formula with new formulas in the theory - SPLITC
<rule name> <position> -> <position>,<position>,...,<position>
\end{lstlisting}
\end{figure}

\subsection{\ourgrounder Grounder}
The \ourgrounder grounder is a proof-of-concept implementation that demonstrates the feasibility of certifying the grounding process, serving as a testbed for refining the proof format and supported rewrite rules defined in \cref{sec:proof}.
It takes as input a FOX problem $\FOXProbshort$ with $\theory$ in GNF and produces a CNF formula $\theory'$ that is $\inputStructure$-equivalent to $\theory$, along with a \ourproofformat proof certifying the grounding.

After parsing the input using a custom \texttt{ANTLR4} parser and internalizing the theory as a vector of abstract syntax trees, the grounder processes each sentence in order.
Each sentence in GNF is a (possibly nested) universal quantifier over a body in Grounding Disjunctive Form.
The grounder applies \iq to instantiate the outermost universal quantifier, restricting the domain to elements satisfying the guard.
Next, it applies \splitc to split the resulting conjunction into one top-level formula per domain element.
Nested universal quantifiers in the body are handled the same way recursively.

Each instantiated formula is traversed inside-out.
These formulas are now all in Grounding Disjunctive Form, so the body is a disjunction of grounding literals, each of which is either a (possibly negated) ground atom or an existentially quantified subformula.
First, existential quantifiers are instantiated via \iq into a disjunction of ground literals, after which they are processed identically to the other literals.
For each (possibly negated) ground atom, \epred or \eprop is applied to evaluate the now grounded atom against $\inputStructure$ and replace it with $\ltrue$ or $\lfalse$ accordingly; if the atom is negated, \stn is applied to flip the resulting truth value.
Once all literals are resolved, \snor collapses the disjunction, short-circuiting to $\ltrue$ if any literal is true or dropping $\lfalse$ literals.
If at any point a formula simplifies to $\lfalse$, the grounder immediately emits \unsat and terminates without completing the remaining sentences.
If a formula simplifies to $\ltrue$, it is dropped via \trivial.
We illustrate the grounding process with a toy example.
\begin{example}
  The goal is to ground the FOX problem $\FOXProb$ with $\inputVocabulary=\{P/1, Q/1\}$, $\outputVocabulary=\{R/1\}$, $\theory=\{\forall x [P(x)]: R(x) \lor \exists y : Q(y).\}$ and $\inputStructure=\{D=\{1,2,3\}, P=\{2,3\}, Q=\{2\}\}$.
  We illustrate the steps applied by the grounder by walking through the obtained proof in \cref{lst:proof}.
  \iq instantiates the universal quantifier only over elements satisfying the guard $P(x)$.
  After \splitc separates the two formulas, the existential subformula $\exists y: Q(y)$ is instantiated over the full domain by a second \iq application, and \epred evaluates each ground atom against $\inputStructure$.
  Since $Q(2)$ holds, \snor short-circuits the disjunction to $\ltrue$, before being dropped via \trivial.
  For the remaining formula, the same steps apply.
  The grounded theory is therefore empty, certified by \certifoxinline{FINAL IDS : -}.

  \begin{figure}[t]
  \begin{lstlisting}[language=certifox, caption={An illustration of a \ourproofformat proof for grounding. Some steps are omitted for brevity.}, label={lst:proof}]
    IQ @ 1 // (1)=(R(2) | ? y : Q(y)) & (R(3) | ? y : Q(y))
    SPLITC   1 -> 2,3 //(2) = (R(2) | ? y : Q(y)), (3)=(R(3) | ? y : Q(y))
    IQ @ 2[1] // (2)=(R(2) | (Q(1) | Q(2) | Q(3)))
    EPRED @ 2[1,0] // (2)=(R(2) | (false | Q(2) | Q(3)))
    EPRED @ 2[1,1] // (2)=(R(2) | (false | true | Q(3)))
    EPRED @ 2[1,2] // (2)=(R(2) | (false | true | false))
    SNOR @ 2[1] // (2)=(R(2) | true)
    SNOR @ 2 // (2)=true
    TRIVIAL - 2  // delete (2)
    // repeat EPRED, SNOR, TRIVIAL for (3)
    FINAL IDS : - // The grounded theory is empty
  \end{lstlisting}
\end{figure}

\end{example}
\subsection{\ourchecker Checker}
The \ourchecker checker is a proof-of-concept implementation that demonstrates the feasibility of independently verifying the correctness of the grounding process using the generated \ourproofformat proofs.
It takes as input the original FOX problem $\FOXProbshort$, the grounded CNF formula $\theory'$ produced by the grounder, and the \ourproofformat proof emitted by the grounder.
It validates the correctness of the grounding process by replaying each logged transformation step and checking that the final theory is syntactically identical to the claimed grounding.

Given that all rules of the \ourtoolkit proof system preserve $\inputStructure$-equivalence, if the checker successfully verifies the proof, it guarantees that the original theory $\theory$ and the grounded theory $\theory'$ are $\inputStructure$-equivalent, which is the key correctness guarantee of \ourtoolkit.
The checker replays the proof sequentially: for each step, it identifies the target subformula by position, applies the specified rule, and rejects immediately if the rule cannot be applied.
If all steps succeed, it verifies that the resulting theory is syntactically identical to the claimed grounding, accepting if they match and rejecting otherwise.

Each rule application requires at most a single traversal of the targeted subformula tree, so the checker runs in time linear in the size of the proof (and the formula). Crucially, this means the checker can be kept intentionally simple and small enough that formal verification becomes feasible, which is exactly what makes it a meaningful trust anchor.

\section{Experimental Validation}

\begin{figure}[t]
\centering
\includegraphics[width=\textwidth]{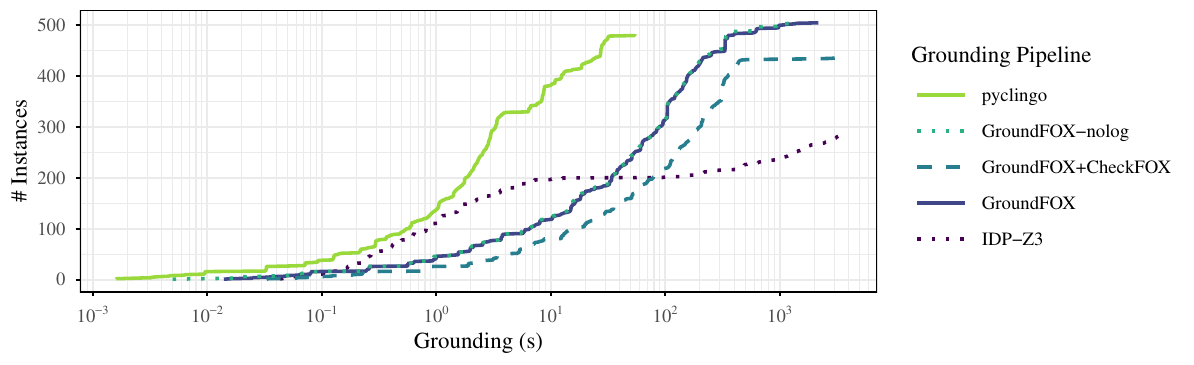}
\caption{Number of instances $y$ with individual completion time $\leq x$ across grounding pipelines.}
\label{fig:groundingruntimes}
\end{figure}

\begin{figure}[t]
\centering
\includegraphics[width=.74\textwidth]{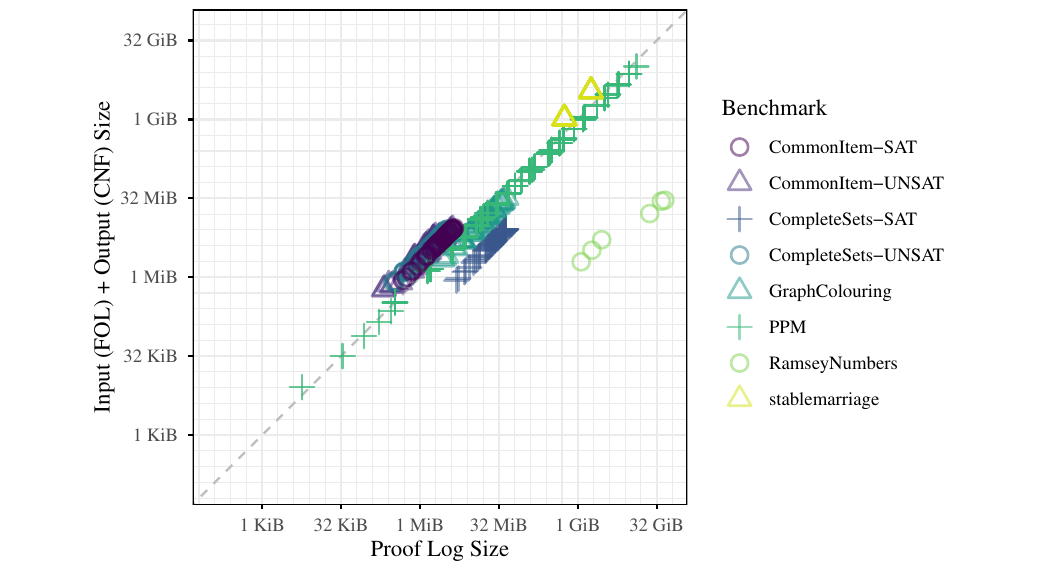}
\caption{Scatter plot of proof sizes for the different benchmarks.}
\label{fig:proofsizes}
\end{figure}

We use the DIRT benchmark suite~\cite{LVV25DIRTLiterature-BasedBenchmarkSuiteGrounders} and select all benchmarks that can be translated into GNF without introducing auxiliary predicates, allowing a fair comparison between the different grounders involved.
Concretely, the eight problem sets we use are (1) CommonItem-SAT, (2) CommonItem-UNSAT, (3) CompleteSets-SAT, (4) CompleteSets-UNSAT, (5) GraphColouring, (6) PPM, (7) RamseyNumbers, and (8) stablemarriage.
For all of these benchmarks, we created a GNF encoding by directly translating (with minimal changes) an existing (IDP) encoding to our syntax.
For one of the problems (stablemarriage), there were multiple IDP encodings; here, we used the encoding on which \idpztool was the most efficient.

\subsection{Setup}
Because of time constraints, we could not compare against all available grounders.
Instead, we compare only against \idpztool (0.12.0) and \pyclingo (5.8.0), which are state-of-the-art solving frameworks for problems expressed in \fodot and ASP, respectively.
These frameworks follow a ground-and-solve approach and therefore contain highly optimized grounders.
We consider three versions of our tool:\footnote{The source code is available at \url{https://gitlab.kuleuven.be/krr/software/groundfox-checkfox}} the grounder (\ourgrounder~0.1.0) without proof logging, the grounder with proof logging, the grounder with proof logging and checking (\ourchecker~0.1.0).

Experiments were conducted on an Intel Xeon Platinum 8468 CPU cluster running Rocky Linux~8.10. Each run was executed as a single-core job with 8 GiB RAM, with multiple runs scheduled concurrently across the cluster.
Each instance was grounded and then solved with a 1-hour time limit.

\subsection{Runtime Results}
First, we compare the runtime of the grounders.
\cref{fig:groundingruntimes} shows grounding runtimes for the benchmark instances across grounding pipelines.
Compared to \pyclingo and \idpztool, the performance of \ourgrounder grounder is respectable overall, although it is outperformed by \pyclingo on all benchmarks and by \idpztool on about half.
\idpztool plateaus after solving around 200 instances: it performs particularly well on the PPM benchmark but struggles with the remaining benchmarks.
We believe this is because the PPM instances involve a small typed integer domain with arithmetic constraints that the underlying Z3 SMT solver handles efficiently.

The plot also shows that the overhead of proof logging is minimal in \ourgrounder, and the overhead of checking is within a factor of 2--3 in most cases.
Interestingly, \ourgrounder was able to ground most instances given the memory and time limits.
In particular, \pyclingo runs out of memory on 30 instances of PPM and runs out of time on 2 instances of stablemarriage, while \ourgrounder grounds all stablemarriage instances successfully and runs out of memory on only 4 instances of PPM\@.

We note that the tools are not equally affected by the memory and time constraints.
As mentioned, out of a total of 515 instances, \pyclingo hit the time limit on 2 and the memory limit on 30, while \idpztool timed out on 117 and ran out of memory on 112 instances.
\ourgrounder (with and without proof logging) timed out on 6 and ran out of memory on only 4 instances.
\ourchecker timed out on 3 instances and ran out of memory on 66 of the 505 successfully grounded instances, making it the most memory-intensive step; addressing this remains future work.

\subsection{Proof Size Results}

We are also interested in the size of the generated proofs, as this is an important factor for the feasibility of proof checking.
\cref{fig:proofsizes} presents a scatter plot of the proof sizes for the different benchmarks.
This figure shows that the proof sizes vary widely across benchmarks, with some proofs being relatively small (a few hundred lines) and others being quite large (millions of lines).
This variation in proof size makes sense given the large variation in domains.
When comparing the proof size against the size of the input and obtained output, we see that the RamseyNumbers proofs are larger.
In the GNF encoding for the RamseyNumbers problems, no guards are used for the quantifiers, which in combination with the deeply nested quantifiers leads to an exponential number of formulas to simplify in the proof.
This stands in contrast to the other problem families: IDP is a typed system supporting multi-sorted logic, and types in an IDP encoding translate directly into guards on the quantifiers in the corresponding GNF encoding.
RamseyNumbers is the only problem family whose IDP encoding does not use types, and therefore yields unguarded quantifiers in the GNF encoding.

\section{Conclusion}
We presented \ourtoolkit, a certifying framework for grounding FOX problems, comprising \ourgrounder, a certifying grounder for \ourNF problems producing compact groundings via domain knowledge; the \ourproofformat proof format; and \ourchecker, an independent proof checker.
Experiments show that \ourgrounder is broadly comparable to \idpztool and \pyclingo, with negligible proof-logging overhead, while \ourchecker verifies proofs within a constant factor of grounding time.
Future work includes broader language coverage (arbitrary FOL sentences, cardinality constraints), a formally verified checker, and seamless integration into high-level solving pipelines, as well as extending the proof system with more complex reasoning.

\newpage
\bibliographystyle{eptcs}
\bibliography{bb-includes/bb-refs, other-refs, bb-includes/bb-refs-to-appear}

\begin{thebibliography}{10}
\providecommand{\bibitemdeclare}[2]{}
\providecommand{\surnamestart}{}
\providecommand{\surnameend}{}
\providecommand{\urlprefix}{Available at }
\providecommand{\url}[1]{\texttt{#1}}
\providecommand{\href}[2]{\texttt{#2}}
\providecommand{\urlalt}[2]{\href{#1}{#2}}
\providecommand{\doi}[1]{doi:\urlalt{https://doi.org/#1}{#1}}
\providecommand{\eprint}[1]{arXiv:\urlalt{https://arxiv.org/abs/#1}{#1}}
\providecommand{\bibinfo}[2]{#2}

\bibitemdeclare{inproceedings}{AGJMN18MetamorphicTestingConstraintSolvers}
\bibitem{AGJMN18MetamorphicTestingConstraintSolvers}
\bibinfo{author}{{\"{O}}zg{\"{u}}r \surnamestart Akg{\"{u}}n\surnameend}, \bibinfo{author}{Ian~P. \surnamestart Gent\surnameend}, \bibinfo{author}{Christopher \surnamestart Jefferson\surnameend}, \bibinfo{author}{Ian \surnamestart Miguel\surnameend} \& \bibinfo{author}{Peter \surnamestart Nightingale\surnameend} (\bibinfo{year}{2018}): \emph{\bibinfo{title}{Metamorphic Testing of Constraint Solvers}}.
\newblock In \bibinfo{editor}{John~N. \surnamestart Hooker\surnameend}, editor: {\slshape \bibinfo{booktitle}{Principles and Practice of Constraint Programming - 24th International Conference, {CP} 2018, Lille, France, August 27-31, 2018, Proceedings}}, {\slshape \bibinfo{series}{Lecture Notes in Computer Science}} \bibinfo{volume}{11008}, \bibinfo{publisher}{Springer}, pp. \bibinfo{pages}{727--736}, \doi{10.1007/978-3-319-98334-9_46}.

\bibitemdeclare{article}{ABMRS11IntroductionCertifyingAlgorithms}
\bibitem{ABMRS11IntroductionCertifyingAlgorithms}
\bibinfo{author}{Eyad \surnamestart Alkassar\surnameend}, \bibinfo{author}{Sascha \surnamestart B{\"{o}}hme\surnameend}, \bibinfo{author}{Kurt \surnamestart Mehlhorn\surnameend}, \bibinfo{author}{Christine \surnamestart Rizkallah\surnameend} \& \bibinfo{author}{Pascal \surnamestart Schweitzer\surnameend} (\bibinfo{year}{2011}): \emph{\bibinfo{title}{An Introduction to Certifying Algorithms}}.
\newblock {\slshape \bibinfo{journal}{it Inf. Technol.}} \bibinfo{volume}{53}(\bibinfo{number}{6}), pp. \bibinfo{pages}{287--293}, \doi{10.1524/itit.2011.0655}.

\bibitemdeclare{article}{ADFHPR19InconsistencyProofsASPASP-DRUPE}
\bibitem{ADFHPR19InconsistencyProofsASPASP-DRUPE}
\bibinfo{author}{Mario \surnamestart Alviano\surnameend}, \bibinfo{author}{Carmine \surnamestart Dodaro\surnameend}, \bibinfo{author}{Johannes~Klaus \surnamestart Fichte\surnameend}, \bibinfo{author}{Markus \surnamestart Hecher\surnameend}, \bibinfo{author}{Tobias \surnamestart Philipp\surnameend} \& \bibinfo{author}{Jakob \surnamestart Rath\surnameend} (\bibinfo{year}{2019}): \emph{\bibinfo{title}{Inconsistency Proofs for {{ASP}:} The {ASP} - {DRUPE} Format}}.
\newblock {\slshape \bibinfo{journal}{Theory Pract. Log. Program.}} \bibinfo{volume}{19}(\bibinfo{number}{5-6}), pp. \bibinfo{pages}{891--907}, \doi{10.1017/S1471068419000255}.

\bibitemdeclare{article}{BBFF20ScalableFine-GrainedProofsFormulaProcessing}
\bibitem{BBFF20ScalableFine-GrainedProofsFormulaProcessing}
\bibinfo{author}{Haniel \surnamestart Barbosa\surnameend}, \bibinfo{author}{Jasmin~Christian \surnamestart Blanchette\surnameend}, \bibinfo{author}{Mathias \surnamestart Fleury\surnameend} \& \bibinfo{author}{Pascal \surnamestart Fontaine\surnameend} (\bibinfo{year}{2020}): \emph{\bibinfo{title}{Scalable Fine-Grained Proofs for Formula Processing}}.
\newblock {\slshape \bibinfo{journal}{J. Autom. Reason.}} \bibinfo{volume}{64}(\bibinfo{number}{3}), pp. \bibinfo{pages}{485--510}, \doi{10.1007/s10817-018-09502-y}.

\bibitemdeclare{inproceedings}{BRKLNNOP22FlexibleProofProductionIndustrial-StrengthSMTSolver}
\bibitem{BRKLNNOP22FlexibleProofProductionIndustrial-StrengthSMTSolver}
\bibinfo{author}{Haniel \surnamestart Barbosa\surnameend}, \bibinfo{author}{Andrew \surnamestart Reynolds\surnameend}, \bibinfo{author}{Gereon \surnamestart Kremer\surnameend}, \bibinfo{author}{Hanna \surnamestart Lachnitt\surnameend}, \bibinfo{author}{Aina \surnamestart Niemetz\surnameend}, \bibinfo{author}{Andres \surnamestart N{\"{o}}tzli\surnameend}, \bibinfo{author}{Alex \surnamestart Ozdemir\surnameend}, \bibinfo{author}{Mathias \surnamestart Preiner\surnameend}, \bibinfo{author}{Arjun \surnamestart Viswanathan\surnameend}, \bibinfo{author}{Scott \surnamestart Viteri\surnameend}, \bibinfo{author}{Yoni \surnamestart Zohar\surnameend}, \bibinfo{author}{Cesare \surnamestart Tinelli\surnameend} \& \bibinfo{author}{Clark~W. \surnamestart Barrett\surnameend} (\bibinfo{year}{2022}): \emph{\bibinfo{title}{Flexible Proof Production in an Industrial-Strength {SMT} Solver}}.
\newblock In \bibinfo{editor}{Jasmin \surnamestart Blanchette\surnameend}, \bibinfo{editor}{Laura \surnamestart Kov{\'{a}}cs\surnameend} \& \bibinfo{editor}{Dirk \surnamestart Pattinson\surnameend}, editors: {\slshape \bibinfo{booktitle}{Automated Reasoning - 11th International Joint Conference, {IJCAR} 2022, Haifa, Israel, August 8-10, 2022, Proceedings}}, {\slshape \bibinfo{series}{Lecture Notes in Computer Science}} \bibinfo{volume}{13385}, \bibinfo{publisher}{Springer}, pp. \bibinfo{pages}{15--35}, \doi{10.1007/978-3-031-10769-6_3}.

\bibitemdeclare{inproceedings}{BBNOV23CertifiedCore-GuidedMaxSATSolving}
\bibitem{BBNOV23CertifiedCore-GuidedMaxSATSolving}
\bibinfo{author}{Jeremias \surnamestart Berg\surnameend}, \bibinfo{author}{Bart \surnamestart Bogaerts\surnameend}, \bibinfo{author}{Jakob \surnamestart Nordstr{\"{o}}m\surnameend}, \bibinfo{author}{Andy \surnamestart Oertel\surnameend} \& \bibinfo{author}{Dieter \surnamestart Vandesande\surnameend} (\bibinfo{year}{2023}): \emph{\bibinfo{title}{Certified Core-Guided {MaxSAT} Solving}}.
\newblock In \bibinfo{editor}{Brigitte \surnamestart Pientka\surnameend} \& \bibinfo{editor}{Cesare \surnamestart Tinelli\surnameend}, editors: {\slshape \bibinfo{booktitle}{Automated Deduction - {CADE} 29 - 29th International Conference on Automated Deduction, Rome, Italy, July 1-4, 2023, Proceedings}}, {\slshape \bibinfo{series}{Lecture Notes in Computer Science}} \bibinfo{volume}{14132}, \bibinfo{publisher}{Springer}, pp. \bibinfo{pages}{1--22}, \doi{10.1007/978-3-031-38499-8_1}.

\bibitemdeclare{inproceedings}{BFBDG26UsingCertifyingConstraintSolversGeneratingStep-wise}
\bibitem{BFBDG26UsingCertifyingConstraintSolversGeneratingStep-wise}
\bibinfo{author}{Ignace \surnamestart Bleukx\surnameend}, \bibinfo{author}{Maarten \surnamestart Flippo\surnameend}, \bibinfo{author}{Bart \surnamestart Bogaerts\surnameend}, \bibinfo{author}{Emir \surnamestart Demirovic\surnameend} \& \bibinfo{author}{Tias \surnamestart Guns\surnameend} (\bibinfo{year}{2026}): \emph{\bibinfo{title}{Using Certifying Constraint Solvers for Generating Step-wise Explanations}}.
\newblock In \bibinfo{editor}{Koenig} et~al.  \cite{aaai/2026}, pp. \bibinfo{pages}{14192--14200}, \doi{10.1609/AAAI.V40I17.38432}.

\bibitemdeclare{article}{BGMN23CertifiedDominanceSymmetryBreakingCombinatorialOptimisation}
\bibitem{BGMN23CertifiedDominanceSymmetryBreakingCombinatorialOptimisation}
\bibinfo{author}{Bart \surnamestart Bogaerts\surnameend}, \bibinfo{author}{Stephan \surnamestart Gocht\surnameend}, \bibinfo{author}{Ciaran \surnamestart McCreesh\surnameend} \& \bibinfo{author}{Jakob \surnamestart Nordstr{\"{o}}m\surnameend} (\bibinfo{year}{2023}): \emph{\bibinfo{title}{Certified Dominance and Symmetry Breaking for Combinatorial Optimisation}}.
\newblock {\slshape \bibinfo{journal}{J. Artif. Intell. Res.}} \bibinfo{volume}{77}, pp. \bibinfo{pages}{1539--1589}, \doi{10.1613/jair.1.14296}.

\bibitemdeclare{inproceedings}{BB09FuzzingDelta-DebuggingSMTSolvers}
\bibitem{BB09FuzzingDelta-DebuggingSMTSolvers}
\bibinfo{author}{Robert \surnamestart Brummayer\surnameend} \& \bibinfo{author}{Armin \surnamestart Biere\surnameend} (\bibinfo{year}{2009}): \emph{\bibinfo{title}{Fuzzing and Delta-Debugging {SMT} Solvers}}.
\newblock In \bibinfo{editor}{Ofer~Strichman \surnamestart Bruno~Dutertre\surnameend}, editor: {\slshape \bibinfo{booktitle}{Proceedings of the 7th International Workshop on Satisfiability Modulo Theories}}, \bibinfo{series}{SMT '09}, \bibinfo{publisher}{Association for Computing Machinery}, \bibinfo{address}{New York, NY, USA}, p. \bibinfo{pages}{1–5}, \doi{10.1145/1670412.1670413}.

\bibitemdeclare{inproceedings}{BLB10AutomatedTestingDebuggingSATQBFSolvers}
\bibitem{BLB10AutomatedTestingDebuggingSATQBFSolvers}
\bibinfo{author}{Robert \surnamestart Brummayer\surnameend}, \bibinfo{author}{Florian \surnamestart Lonsing\surnameend} \& \bibinfo{author}{Armin \surnamestart Biere\surnameend} (\bibinfo{year}{2010}): \emph{\bibinfo{title}{Automated Testing and Debugging of {SAT} and {QBF} Solvers}}.
\newblock In \bibinfo{editor}{Ofer \surnamestart Strichman\surnameend} \& \bibinfo{editor}{Stefan \surnamestart Szeider\surnameend}, editors: {\slshape \bibinfo{booktitle}{Theory and Applications of Satisfiability Testing - {SAT} 2010, 13th International Conference, {SAT} 2010, Edinburgh, {UK}, July 11-14, 2010. Proceedings}}, {\slshape \bibinfo{series}{Lecture Notes in Computer Science}} \bibinfo{volume}{6175}, \bibinfo{publisher}{Springer}, pp. \bibinfo{pages}{44--57}, \doi{10.1007/978-3-642-14186-7_6}.

\bibitemdeclare{article}{CVVD22IDP-Z3reasoningengineFO}
\bibitem{CVVD22IDP-Z3reasoningengineFO}
\bibinfo{author}{Pierre \surnamestart Carbonnelle\surnameend}, \bibinfo{author}{Simon \surnamestart Vandevelde\surnameend}, \bibinfo{author}{Joost \surnamestart Vennekens\surnameend} \& \bibinfo{author}{Marc \surnamestart Denecker\surnameend} (\bibinfo{year}{2022}): \emph{\bibinfo{title}{{IDP-{Z3}:} a reasoning engine for {{FO($\cdot$)}}}}.
\newblock {\slshape \bibinfo{journal}{CoRR}} \bibinfo{volume}{abs/2202.00343}, \doi{10.48550/arXiv.2202.00343}.
\newblock \eprint{2202.00343}.

\bibitemdeclare{article}{CKSW13hybridbranch-and-boundapproachexactrationalmixed-integer}
\bibitem{CKSW13hybridbranch-and-boundapproachexactrationalmixed-integer}
\bibinfo{author}{William~J. \surnamestart Cook\surnameend}, \bibinfo{author}{Thorsten \surnamestart Koch\surnameend}, \bibinfo{author}{Daniel~E. \surnamestart Steffy\surnameend} \& \bibinfo{author}{Kati \surnamestart Wolter\surnameend} (\bibinfo{year}{2013}): \emph{\bibinfo{title}{A hybrid branch-and-bound approach for exact rational mixed-integer programming}}.
\newblock {\slshape \bibinfo{journal}{Math. Program. Comput.}} \bibinfo{volume}{5}(\bibinfo{number}{3}), pp. \bibinfo{pages}{305--344}, \doi{10.1007/s12532-013-0055-6}.

\bibitemdeclare{inproceedings}{DeharbeFP11}
\bibitem{DeharbeFP11}
\bibinfo{author}{David \surnamestart D{\'{e}}harbe\surnameend}, \bibinfo{author}{Pascal \surnamestart Fontaine\surnameend} \& \bibinfo{author}{Bruno~Woltzenlogel \surnamestart Paleo\surnameend} (\bibinfo{year}{2011}): \emph{\bibinfo{title}{Quantifier Inference Rules for {SMT} proofs}}.
\newblock In \bibinfo{editor}{Pascal \surnamestart Fontaine\surnameend} \& \bibinfo{editor}{Aaron \surnamestart Stump\surnameend}, editors: {\slshape \bibinfo{booktitle}{PxTP 2011: First International Workshop on Proof eXchange for Theorem Proving, Wroc{\l}aw, Poland, August 1, 2011}}, pp. \bibinfo{pages}{33--39}.
\newblock \urlprefix\url{https://inria.hal.science/hal-00642535}.

\bibitemdeclare{inproceedings}{ERH17UnsolvabilityCertificatesClassicalPlanning}
\bibitem{ERH17UnsolvabilityCertificatesClassicalPlanning}
\bibinfo{author}{Salom{\'{e}} \surnamestart Eriksson\surnameend}, \bibinfo{author}{Gabriele \surnamestart R{\"{o}}ger\surnameend} \& \bibinfo{author}{Malte \surnamestart Helmert\surnameend} (\bibinfo{year}{2017}): \emph{\bibinfo{title}{Unsolvability Certificates for Classical Planning}}.
\newblock In \bibinfo{editor}{Laura \surnamestart Barbulescu\surnameend}, \bibinfo{editor}{Jeremy \surnamestart Frank\surnameend}, \bibinfo{editor}{\surnamestart Mausam\surnameend} \& \bibinfo{editor}{Stephen~F. \surnamestart Smith\surnameend}, editors: {\slshape \bibinfo{booktitle}{Proceedings of the Twenty-Seventh International Conference on Automated Planning and Scheduling, {ICAPS} 2017, Pittsburgh, Pennsylvania, {USA}, June 18-23, 2017}}, \bibinfo{publisher}{{AAAI} Press}, pp. \bibinfo{pages}{88--97}, \doi{10.1609/icaps.v27i1.13818}.
\newblock \urlprefix\url{https://aaai.org/ocs/index.php/ICAPS/ICAPS17/paper/view/15734}.

\bibitemdeclare{inproceedings}{FSMSD24Multi-StageProofLoggingFrameworkCertifyCorrectness}
\bibitem{FSMSD24Multi-StageProofLoggingFrameworkCertifyCorrectness}
\bibinfo{author}{Maarten \surnamestart Flippo\surnameend}, \bibinfo{author}{Konstantin \surnamestart Sidorov\surnameend}, \bibinfo{author}{Imko \surnamestart Marijnissen\surnameend}, \bibinfo{author}{Jeff \surnamestart Smits\surnameend} \& \bibinfo{author}{Emir \surnamestart Demirovic\surnameend} (\bibinfo{year}{2024}): \emph{\bibinfo{title}{A Multi-Stage Proof Logging Framework to Certify the Correctness of {CP} Solvers}}.
\newblock In \bibinfo{editor}{Paul \surnamestart Shaw\surnameend}, editor: {\slshape \bibinfo{booktitle}{30th International Conference on Principles and Practice of Constraint Programming, {CP} 2024, September 2-6, 2024, Girona, Spain}}, {\slshape \bibinfo{series}{LIPIcs}} \bibinfo{volume}{307}, \bibinfo{publisher}{Schloss Dagstuhl - Leibniz-Zentrum f{\"{u}}r Informatik}, pp. \bibinfo{pages}{11:1--11:20}, \doi{10.4230/LIPICS.CP.2024.11}.

\bibitemdeclare{inproceedings}{GKKOSW16TheorySolvingMadeEasyClingo5}
\bibitem{GKKOSW16TheorySolvingMadeEasyClingo5}
\bibinfo{author}{Martin \surnamestart Gebser\surnameend}, \bibinfo{author}{Roland \surnamestart Kaminski\surnameend}, \bibinfo{author}{Benjamin \surnamestart Kaufmann\surnameend}, \bibinfo{author}{Max \surnamestart Ostrowski\surnameend}, \bibinfo{author}{Torsten \surnamestart Schaub\surnameend} \& \bibinfo{author}{Philipp \surnamestart Wanko\surnameend} (\bibinfo{year}{2016}): \emph{\bibinfo{title}{Theory Solving Made Easy with Clingo 5}}.
\newblock In \bibinfo{editor}{Manuel \surnamestart Carro\surnameend}, \bibinfo{editor}{Andy \surnamestart King\surnameend}, \bibinfo{editor}{Neda \surnamestart Saeedloei\surnameend} \& \bibinfo{editor}{Marina~De \surnamestart Vos\surnameend}, editors: {\slshape \bibinfo{booktitle}{Technical Communications of the 32nd International Conference on Logic Programming, {ICLP} 2016 TCs, October 16-21, 2016, New York City, {USA}}}, {\slshape \bibinfo{series}{OASIcs}}~\bibinfo{volume}{52}, \bibinfo{publisher}{Schloss Dagstuhl - Leibniz-Zentrum f{\"{u}}r Informatik}, pp. \bibinfo{pages}{2:1--2:15}, \doi{10.4230/OASIcs.ICLP.2016.2}.

\bibitemdeclare{inproceedings}{GST07GrinGoNewGrounderAnswerSetProgramming}
\bibitem{GST07GrinGoNewGrounderAnswerSetProgramming}
\bibinfo{author}{Martin \surnamestart Gebser\surnameend}, \bibinfo{author}{Torsten \surnamestart Schaub\surnameend} \& \bibinfo{author}{Sven \surnamestart Thiele\surnameend} (\bibinfo{year}{2007}): \emph{\bibinfo{title}{GrinGo : {A} New Grounder for Answer Set Programming}}.
\newblock In \bibinfo{editor}{Chitta \surnamestart Baral\surnameend}, \bibinfo{editor}{Gerhard \surnamestart Brewka\surnameend} \& \bibinfo{editor}{John~S. \surnamestart Schlipf\surnameend}, editors: {\slshape \bibinfo{booktitle}{Logic Programming and Nonmonotonic Reasoning, 9th International Conference, {LPNMR} 2007, Tempe, AZ, {USA}, May 15-17, 2007, Proceedings}}, {\slshape \bibinfo{series}{Lecture Notes in Computer Science}} \bibinfo{volume}{4483}, \bibinfo{publisher}{Springer}, pp. \bibinfo{pages}{266--271}, \doi{10.1007/978-3-540-72200-7_24}.

\bibitemdeclare{inproceedings}{GSD19SolverCheckDeclarativeTestingConstraints}
\bibitem{GSD19SolverCheckDeclarativeTestingConstraints}
\bibinfo{author}{Xavier \surnamestart Gillard\surnameend}, \bibinfo{author}{Pierre \surnamestart Schaus\surnameend} \& \bibinfo{author}{Yves \surnamestart Deville\surnameend} (\bibinfo{year}{2019}): \emph{\bibinfo{title}{SolverCheck: Declarative Testing of Constraints}}.
\newblock In \bibinfo{editor}{Thomas \surnamestart Schiex\surnameend} \& \bibinfo{editor}{Simon \surnamestart de~Givry\surnameend}, editors: {\slshape \bibinfo{booktitle}{Principles and Practice of Constraint Programming - 25th International Conference, {CP} 2019, Stamford, CT, {USA}, September 30 - October 4, 2019, Proceedings}}, {\slshape \bibinfo{series}{Lecture Notes in Computer Science}} \bibinfo{volume}{11802}, \bibinfo{publisher}{Springer}, pp. \bibinfo{pages}{565--582}, \doi{10.1007/978-3-030-30048-7_33}.

\bibitemdeclare{inproceedings}{GMNO22CertifiedCNFTranslationsPseudo-BooleanSolving}
\bibitem{GMNO22CertifiedCNFTranslationsPseudo-BooleanSolving}
\bibinfo{author}{Stephan \surnamestart Gocht\surnameend}, \bibinfo{author}{Ruben \surnamestart Martins\surnameend}, \bibinfo{author}{Jakob \surnamestart Nordstr{\"{o}}m\surnameend} \& \bibinfo{author}{Andy \surnamestart Oertel\surnameend} (\bibinfo{year}{2022}): \emph{\bibinfo{title}{Certified {CNF} Translations for Pseudo-{Boolean} Solving}}.
\newblock In \bibinfo{editor}{Kuldeep~S. \surnamestart Meel\surnameend} \& \bibinfo{editor}{Ofer \surnamestart Strichman\surnameend}, editors: {\slshape \bibinfo{booktitle}{25th International Conference on Theory and Applications of Satisfiability Testing, {SAT} 2022, August 2-5, 2022, Haifa, Israel}}, {\slshape \bibinfo{series}{LIPIcs}} \bibinfo{volume}{236}, \bibinfo{publisher}{Schloss Dagstuhl - Leibniz-Zentrum f{\"{u}}r Informatik}, pp. \bibinfo{pages}{16:1--16:25}, \doi{10.4230/LIPIcs.SAT.2022.16}.

\bibitemdeclare{inproceedings}{GMN22AuditableConstraintProgrammingSolver}
\bibitem{GMN22AuditableConstraintProgrammingSolver}
\bibinfo{author}{Stephan \surnamestart Gocht\surnameend}, \bibinfo{author}{Ciaran \surnamestart McCreesh\surnameend} \& \bibinfo{author}{Jakob \surnamestart Nordstr{\"{o}}m\surnameend} (\bibinfo{year}{2022}): \emph{\bibinfo{title}{An Auditable Constraint Programming Solver}}.
\newblock In \bibinfo{editor}{Christine \surnamestart Solnon\surnameend}, editor: {\slshape \bibinfo{booktitle}{28th International Conference on Principles and Practice of Constraint Programming, {CP} 2022, July 31 to August 8, 2022, Haifa, Israel}}, {\slshape \bibinfo{series}{LIPIcs}} \bibinfo{volume}{235}, \bibinfo{publisher}{Schloss Dagstuhl - Leibniz-Zentrum f{\"{u}}r Informatik}, pp. \bibinfo{pages}{25:1--25:18}, \doi{10.4230/LIPIcs.CP.2022.25}.

\bibitemdeclare{inproceedings}{HHW13Trimmingwhilecheckingclausalproofs}
\bibitem{HHW13Trimmingwhilecheckingclausalproofs}
\bibinfo{author}{Marijn \surnamestart Heule\surnameend}, \bibinfo{author}{Warren~A. \surnamestart Hunt\surnameend, Jr.} \& \bibinfo{author}{Nathan \surnamestart Wetzler\surnameend} (\bibinfo{year}{2013}): \emph{\bibinfo{title}{Trimming while checking clausal proofs}}.
\newblock In: {\slshape \bibinfo{booktitle}{Formal Methods in Computer-Aided Design, {FMCAD} 2013, Portland, OR, {USA}, October 20-23, 2013}}, \bibinfo{publisher}{{IEEE}}, pp. \bibinfo{pages}{181--188}, \doi{10.1109/FMCAD.2013.6679408}.
\newblock \urlprefix\url{https://ieeexplore.ieee.org/document/6679408/}.

\bibitemdeclare{inproceedings}{HSB14UnifiedProofSystemQBFPreprocessing}
\bibitem{HSB14UnifiedProofSystemQBFPreprocessing}
\bibinfo{author}{Marijn \surnamestart Heule\surnameend}, \bibinfo{author}{Martina \surnamestart Seidl\surnameend} \& \bibinfo{author}{Armin \surnamestart Biere\surnameend} (\bibinfo{year}{2014}): \emph{\bibinfo{title}{A Unified Proof System for {QBF} Preprocessing}}.
\newblock In \bibinfo{editor}{St{\'{e}}phane \surnamestart Demri\surnameend}, \bibinfo{editor}{Deepak \surnamestart Kapur\surnameend} \& \bibinfo{editor}{Christoph \surnamestart Weidenbach\surnameend}, editors: {\slshape \bibinfo{booktitle}{Automated Reasoning - 7th International Joint Conference, {IJCAR} 2014, Held as Part of the Vienna Summer of Logic, {VSL} 2014, Vienna, Austria, July 19-22, 2014. Proceedings}}, {\slshape \bibinfo{series}{Lecture Notes in Computer Science}} \bibinfo{volume}{8562}, \bibinfo{publisher}{Springer}, pp. \bibinfo{pages}{91--106}, \doi{10.1007/978-3-319-08587-6_7}.

\bibitemdeclare{inproceedings}{Hitarth2024}
\bibitem{Hitarth2024}
\bibinfo{author}{S.~\surnamestart Hitarth\surnameend}, \bibinfo{author}{Cayden~R. \surnamestart Codel\surnameend}, \bibinfo{author}{Hanna \surnamestart Lachnitt\surnameend} \& \bibinfo{author}{Bruno \surnamestart Dutertre\surnameend} (\bibinfo{year}{2024}): \emph{\bibinfo{title}{Extending {DRAT} to {SMT}}}.
\newblock In \bibinfo{editor}{Nina \surnamestart Narodytska\surnameend} \& \bibinfo{editor}{Philipp \surnamestart R{\"{u}}mmer\surnameend}, editors: {\slshape \bibinfo{booktitle}{Formal Methods in Computer-Aided Design, {FMCAD} 2024, Prague, Czech Republic, October 15-18, 2024}}, \bibinfo{publisher}{{IEEE}}, pp. \bibinfo{pages}{1--11}, \doi{10.34727/2024/ISBN.978-3-85448-065-5\_8}.

\bibitemdeclare{inproceedings}{IOTBJMN24CertifiedMaxSATPreprocessing}
\bibitem{IOTBJMN24CertifiedMaxSATPreprocessing}
\bibinfo{author}{Hannes \surnamestart Ihalainen\surnameend}, \bibinfo{author}{Andy \surnamestart Oertel\surnameend}, \bibinfo{author}{Yong~Kiam \surnamestart Tan\surnameend}, \bibinfo{author}{Jeremias \surnamestart Berg\surnameend}, \bibinfo{author}{Matti \surnamestart J{\"{a}}rvisalo\surnameend}, \bibinfo{author}{Magnus~O. \surnamestart Myreen\surnameend} \& \bibinfo{author}{Jakob \surnamestart Nordstr{\"{o}}m\surnameend} (\bibinfo{year}{2024}): \emph{\bibinfo{title}{Certified {MaxSAT} Preprocessing}}.
\newblock In \bibinfo{editor}{Christoph \surnamestart Benzm{\"{u}}ller\surnameend}, \bibinfo{editor}{Marijn J.~H. \surnamestart Heule\surnameend} \& \bibinfo{editor}{Renate~A. \surnamestart Schmidt\surnameend}, editors: {\slshape \bibinfo{booktitle}{Automated Reasoning - 12th International Joint Conference, {IJCAR} 2024, Nancy, France, July 3-6, 2024, Proceedings, Part {I}}}, {\slshape \bibinfo{series}{Lecture Notes in Computer Science}} \bibinfo{volume}{14739}, \bibinfo{publisher}{Springer}, pp. \bibinfo{pages}{396--418}, \doi{10.1007/978-3-031-63498-7_24}.

\bibitemdeclare{inproceedings}{IVSBBJ26EfficientReliableHitting-SetComputationsImplicitHitting}
\bibitem{IVSBBJ26EfficientReliableHitting-SetComputationsImplicitHitting}
\bibinfo{author}{Hannes \surnamestart Ihalainen\surnameend}, \bibinfo{author}{Dieter \surnamestart Vandesande\surnameend}, \bibinfo{author}{Andr{\'{e}} \surnamestart Schidler\surnameend}, \bibinfo{author}{Jeremias \surnamestart Berg\surnameend}, \bibinfo{author}{Bart \surnamestart Bogaerts\surnameend} \& \bibinfo{author}{Matti \surnamestart J{\"{a}}rvisalo\surnameend} (\bibinfo{year}{2026}): \emph{\bibinfo{title}{Efficient and Reliable Hitting-Set Computations for the Implicit Hitting Set Approach}}.
\newblock In \bibinfo{editor}{Koenig} et~al.  \cite{aaai/2026}, pp. \bibinfo{pages}{14251--14260}, \doi{10.1609/AAAI.V40I17.38439}.

\bibitemdeclare{inproceedings}{JBBJ25CertifyingParetoOptimalityMulti-ObjectiveMaximumSatisfiability}
\bibitem{JBBJ25CertifyingParetoOptimalityMulti-ObjectiveMaximumSatisfiability}
\bibinfo{author}{Christoph \surnamestart Jabs\surnameend}, \bibinfo{author}{Jeremias \surnamestart Berg\surnameend}, \bibinfo{author}{Bart \surnamestart Bogaerts\surnameend} \& \bibinfo{author}{Matti \surnamestart J{\"{a}}rvisalo\surnameend} (\bibinfo{year}{2025}): \emph{\bibinfo{title}{Certifying Pareto Optimality in Multi-Objective Maximum Satisfiability}}.
\newblock In \bibinfo{editor}{Arie \surnamestart Gurfinkel\surnameend} \& \bibinfo{editor}{Marijn \surnamestart Heule\surnameend}, editors: {\slshape \bibinfo{booktitle}{Tools and Algorithms for the Construction and Analysis of Systems - 31st International Conference, {TACAS} 2025, Held as Part of the International Joint Conferences on Theory and Practice of Software, {ETAPS} 2025, Hamilton, ON, Canada, May 3-8, 2025, Proceedings, Part {II}}}, {\slshape \bibinfo{series}{Lecture Notes in Computer Science}} \bibinfo{volume}{15697}, \bibinfo{publisher}{Springer}, pp. \bibinfo{pages}{108--129}, \doi{10.1007/978-3-031-90653-4_6}.

\bibitemdeclare{inproceedings}{JHB12InprocessingRules}
\bibitem{JHB12InprocessingRules}
\bibinfo{author}{Matti \surnamestart J{\"{a}}rvisalo\surnameend}, \bibinfo{author}{Marijn \surnamestart Heule\surnameend} \& \bibinfo{author}{Armin \surnamestart Biere\surnameend} (\bibinfo{year}{2012}): \emph{\bibinfo{title}{Inprocessing Rules}}.
\newblock In \bibinfo{editor}{Bernhard \surnamestart Gramlich\surnameend}, \bibinfo{editor}{Dale \surnamestart Miller\surnameend} \& \bibinfo{editor}{Uli \surnamestart Sattler\surnameend}, editors: {\slshape \bibinfo{booktitle}{Automated Reasoning - 6th International Joint Conference, {IJCAR} 2012, Manchester, {UK}, June 26-29, 2012. Proceedings}}, {\slshape \bibinfo{series}{Lecture Notes in Computer Science}} \bibinfo{volume}{7364}, \bibinfo{publisher}{Springer}, pp. \bibinfo{pages}{355--370}, \doi{10.1007/978-3-642-31365-3_28}.

\bibitemdeclare{proceedings}{aaai/2026}
\bibitem{aaai/2026}
\bibinfo{editor}{Sven \surnamestart Koenig\surnameend}, \bibinfo{editor}{Chad \surnamestart Jenkins\surnameend} \& \bibinfo{editor}{Matthew~E. \surnamestart Taylor\surnameend}, editors (\bibinfo{year}{2026}): \emph{\bibinfo{title}{Fortieth {AAAI} Conference on Artificial Intelligence, Thirty-Eighth Conference on Innovative Applications of Artificial Intelligence, Sixteenth Symposium on Educational Advances in Artificial Intelligence, {AAAI} 2026, Singapore, January 20-27, 2026}}. \bibinfo{publisher}{{AAAI} Press}.
\newblock \urlprefix\url{https://aaai.org/proceeding/aaai-40-2026/}.

\bibitemdeclare{inproceedings}{KLMNOTV25PracticallyFeasibleProofLoggingPseudo-BooleanOptimization}
\bibitem{KLMNOTV25PracticallyFeasibleProofLoggingPseudo-BooleanOptimization}
\bibinfo{author}{Wietze \surnamestart Koops\surnameend}, \bibinfo{author}{Daniel \surnamestart {Le Berre}\surnameend}, \bibinfo{author}{Magnus~O. \surnamestart Myreen\surnameend}, \bibinfo{author}{Jakob \surnamestart Nordstr{\"{o}}m\surnameend}, \bibinfo{author}{Andy \surnamestart Oertel\surnameend}, \bibinfo{author}{Yong~Kiam \surnamestart Tan\surnameend} \& \bibinfo{author}{Marc \surnamestart Vinyals\surnameend} (\bibinfo{year}{2025}): \emph{\bibinfo{title}{Practically Feasible Proof Logging for Pseudo-{Boolean} Optimization}}.
\newblock In \bibinfo{editor}{Maria \surnamestart {Garcia de la Banda}\surnameend}, editor: {\slshape \bibinfo{booktitle}{31st International Conference on Principles and Practice of Constraint Programming, {CP} 2025, August 10-15, 2025, Glasgow, Scotland}}, {\slshape \bibinfo{series}{LIPIcs}} \bibinfo{volume}{340}, \bibinfo{publisher}{Schloss Dagstuhl - Leibniz-Zentrum f{\"{u}}r Informatik}, pp. \bibinfo{pages}{21:1--21:27}, \doi{10.4230/LIPICS.CP.2025.21}.

\bibitemdeclare{inproceedings}{LVV25DIRTLiterature-BasedBenchmarkSuiteGrounders}
\bibitem{LVV25DIRTLiterature-BasedBenchmarkSuiteGrounders}
\bibinfo{author}{Lucas~Van \surnamestart Laer\surnameend}, \bibinfo{author}{Simon \surnamestart Vandevelde\surnameend} \& \bibinfo{author}{Joost \surnamestart Vennekens\surnameend} (\bibinfo{year}{2025}): \emph{\bibinfo{title}{{DIRT:} a Literature-Based Benchmark Suite for Grounders}}.
\newblock In \bibinfo{editor}{Giovanni \surnamestart Casini\surnameend}, \bibinfo{editor}{Besik \surnamestart Dundua\surnameend} \& \bibinfo{editor}{Temur \surnamestart Kutsia\surnameend}, editors: {\slshape \bibinfo{booktitle}{Logics in Artificial Intelligence - 19th European Conference, {JELIA} 2025, Kutaisi, Georgia, September 1-4, 2025, Proceedings, Part {I}}}, {\slshape \bibinfo{series}{Lecture Notes in Computer Science}} \bibinfo{volume}{16093}, \bibinfo{publisher}{Springer}, pp. \bibinfo{pages}{343--356}, \doi{10.1007/978-3-032-04587-4_21}.

\bibitemdeclare{article}{MO14PairedAltruisticKidneyDonationUKAlgorithms}
\bibitem{MO14PairedAltruisticKidneyDonationUKAlgorithms}
\bibinfo{author}{David~F. \surnamestart Manlove\surnameend} \& \bibinfo{author}{Gregg \surnamestart O'Malley\surnameend} (\bibinfo{year}{2014}): \emph{\bibinfo{title}{Paired and Altruistic Kidney Donation in the {{UK}:} Algorithms and Experimentation}}.
\newblock {\slshape \bibinfo{journal}{{ACM} J. Exp. Algorithmics}} \bibinfo{volume}{19}(\bibinfo{number}{1}), \doi{10.1145/2670129}.

\bibitemdeclare{article}{MMNS11Certifyingalgorithms}
\bibitem{MMNS11Certifyingalgorithms}
\bibinfo{author}{Ross~M. \surnamestart McConnell\surnameend}, \bibinfo{author}{Kurt \surnamestart Mehlhorn\surnameend}, \bibinfo{author}{Stefan \surnamestart N{\"{a}}her\surnameend} \& \bibinfo{author}{Pascal \surnamestart Schweitzer\surnameend} (\bibinfo{year}{2011}): \emph{\bibinfo{title}{Certifying algorithms}}.
\newblock {\slshape \bibinfo{journal}{Comput. Sci. Rev.}} \bibinfo{volume}{5}(\bibinfo{number}{2}), pp. \bibinfo{pages}{119--161}, \doi{10.1016/j.cosrev.2010.09.009}.

\bibitemdeclare{article}{Michael1995}
\bibitem{Michael1995}
\bibinfo{author}{Michaelis \surnamestart Michael\surnameend} \& \bibinfo{author}{A.~V. \surnamestart Townsend\surnameend} (\bibinfo{year}{1995}): \emph{\bibinfo{title}{Binary Quantification Systems}}.
\newblock {\slshape \bibinfo{journal}{Notre Dame Journal of Formal Logic}} \bibinfo{volume}{36}(\bibinfo{number}{3}), pp. \bibinfo{pages}{382--395}, \doi{10.1305/ndjfl/1040149354}.

\bibitemdeclare{inproceedings}{MT05FrameworkRepresentingSolvingNPSearchProblems}
\bibitem{MT05FrameworkRepresentingSolvingNPSearchProblems}
\bibinfo{author}{David~G. \surnamestart Mitchell\surnameend} \& \bibinfo{author}{Eugenia \surnamestart Ternovska\surnameend} (\bibinfo{year}{2005}): \emph{\bibinfo{title}{A Framework for Representing and Solving {NP} Search Problems}}.
\newblock In \bibinfo{editor}{Manuela~M. \surnamestart Veloso\surnameend} \& \bibinfo{editor}{Subbarao \surnamestart Kambhampati\surnameend}, editors: {\slshape \bibinfo{booktitle}{Proceedings, The Twentieth National Conference on Artificial Intelligence and the Seventeenth Innovative Applications of Artificial Intelligence Conference, July 9-13, 2005, Pittsburgh, Pennsylvania, {USA}}}, \bibinfo{publisher}{{AAAI} Press / The {MIT} Press}, pp. \bibinfo{pages}{430--435}.
\newblock \urlprefix\url{http://www.aaai.org/Library/AAAI/2005/aaai05-068.php}.

\bibitemdeclare{inproceedings}{NBGWB01A-PrologDecisionSupportSystemSpaceShuttle}
\bibitem{NBGWB01A-PrologDecisionSupportSystemSpaceShuttle}
\bibinfo{author}{Monica~L. \surnamestart Nogueira\surnameend}, \bibinfo{author}{Marcello \surnamestart Balduccini\surnameend}, \bibinfo{author}{Michael \surnamestart Gelfond\surnameend}, \bibinfo{author}{Richard \surnamestart Watson\surnameend} \& \bibinfo{author}{Matthew \surnamestart Barry\surnameend} (\bibinfo{year}{2001}): \emph{\bibinfo{title}{An A-{Prolog} Decision Support System for the Space Shuttle}}.
\newblock In \bibinfo{editor}{I.~V. \surnamestart Ramakrishnan\surnameend}, editor: {\slshape \bibinfo{booktitle}{Practical Aspects of Declarative Languages, Third International Symposium, {PADL} 2001, Las Vegas, Nevada, {USA}, March 11-12, 2001, Proceedings}}, {\slshape \bibinfo{series}{Lecture Notes in Computer Science}} \bibinfo{volume}{1990}, \bibinfo{publisher}{Springer}, pp. \bibinfo{pages}{169--183}, \doi{10.1007/3-540-45241-9_12}.

\bibitemdeclare{inproceedings}{NBNPRT22ReconstructingFine-GrainedProofsRewritesDSL}
\bibitem{NBNPRT22ReconstructingFine-GrainedProofsRewritesDSL}
\bibinfo{author}{Andres \surnamestart N{\"{o}}tzli\surnameend}, \bibinfo{author}{Haniel \surnamestart Barbosa\surnameend}, \bibinfo{author}{Aina \surnamestart Niemetz\surnameend}, \bibinfo{author}{Mathias \surnamestart Preiner\surnameend}, \bibinfo{author}{Andrew \surnamestart Reynolds\surnameend}, \bibinfo{author}{Clark~W. \surnamestart Barrett\surnameend} \& \bibinfo{author}{Cesare \surnamestart Tinelli\surnameend} (\bibinfo{year}{2022}): \emph{\bibinfo{title}{Reconstructing Fine-Grained Proofs of Rewrites Using a Domain-Specific Language}}.
\newblock In: {\slshape \bibinfo{booktitle}{Formal Methods in Computer-Aided Design (FMCAD)}}, pp. \bibinfo{pages}{65--74}, \doi{10.34727/2022/isbn.978-3-85448-053-2_10}.

\bibitemdeclare{techreport}{S98ImplementationLocalGroundingLogicProgramsStable}
\bibitem{S98ImplementationLocalGroundingLogicProgramsStable}
\bibinfo{author}{Tommi \surnamestart Syrj{\"a}nen\surnameend} (\bibinfo{year}{1998}): \emph{\bibinfo{title}{Implementation of Local Grounding for Logic Programs with Stable Model Semantics}}.
\newblock \bibinfo{type}{Technical Report} \bibinfo{number}{B18}, \bibinfo{institution}{Helsinki University of Technology, Finland}.

\bibitemdeclare{inproceedings}{VCB26CertifiedBranch-and-BoundMaxSATSolving}
\bibitem{VCB26CertifiedBranch-and-BoundMaxSATSolving}
\bibinfo{author}{Dieter \surnamestart Vandesande\surnameend}, \bibinfo{author}{Jordi \surnamestart Coll\surnameend} \& \bibinfo{author}{Bart \surnamestart Bogaerts\surnameend} (\bibinfo{year}{2026}): \emph{\bibinfo{title}{Certified Branch-and-Bound {MaxSAT} Solving}}.
\newblock In \bibinfo{editor}{Koenig} et~al.  \cite{aaai/2026}, pp. \bibinfo{pages}{14342--14351}, \doi{10.1609/AAAI.V40I17.38449}.

\bibitemdeclare{inproceedings}{VDB22QMaxSATpbCertifiedMaxSATSolver}
\bibitem{VDB22QMaxSATpbCertifiedMaxSATSolver}
\bibinfo{author}{Dieter \surnamestart Vandesande\surnameend}, \bibinfo{author}{Wolf \surnamestart {De Wulf}\surnameend} \& \bibinfo{author}{Bart \surnamestart Bogaerts\surnameend} (\bibinfo{year}{2022}): \emph{\bibinfo{title}{{QMaxSATpb}: {A} Certified {MaxSAT} Solver}}.
\newblock In \bibinfo{editor}{Georg \surnamestart Gottlob\surnameend}, \bibinfo{editor}{Daniela \surnamestart Inclezan\surnameend} \& \bibinfo{editor}{Marco \surnamestart Maratea\surnameend}, editors: {\slshape \bibinfo{booktitle}{Logic Programming and Nonmonotonic Reasoning - 16th International Conference, {LPNMR} 2022, Genova, Italy, September 5-9, 2022, Proceedings}}, {\slshape \bibinfo{series}{Lecture Notes in Computer Science}} \bibinfo{volume}{13416}, \bibinfo{publisher}{Springer}, pp. \bibinfo{pages}{429--442}, \doi{10.1007/978-3-031-15707-3_33}.

\bibitemdeclare{article}{WMD10GroundingFOFOIDBounds}
\bibitem{WMD10GroundingFOFOIDBounds}
\bibinfo{author}{Johan \surnamestart Wittocx\surnameend}, \bibinfo{author}{Maarten \surnamestart Mari{\"{e}}n\surnameend} \& \bibinfo{author}{Marc \surnamestart Denecker\surnameend} (\bibinfo{year}{2010}): \emph{\bibinfo{title}{Grounding {FO} and {FO(ID)} with Bounds}}.
\newblock {\slshape \bibinfo{journal}{J. Artif. Intell. Res.}} \bibinfo{volume}{38}, pp. \bibinfo{pages}{223--269}, \doi{10.1613/jair.2980}.

\end{thebibliography}

\end{document}